\documentclass[twocolumn, 10pt]{IEEEtran}
\usepackage{graphicx}
\usepackage{amssymb}
\usepackage{amsmath}
\usepackage{mathtools}
\usepackage{dsfont}
\usepackage{cite}
\usepackage{stfloats}
\usepackage{subfigure}
\usepackage{psfrag}
\usepackage[mathscr]{euscript}
\usepackage{acronym}  
\usepackage{booktabs}
\usepackage{bbm}

\usepackage{float}
\usepackage{balance}

\acrodef{CCDF}{complementary cumulative distribution function}
\acrodef{CF}{characteristic function}
\acrodef{PPP}{Poisson point processe}
\acrodef{RV}{random variable}
\acrodef{i.i.d.}{independent and identically distributed}
\acrodef{PDF}{probability distribution function}
\acrodef{CDF}{cumulative distribution function}
\acrodef{ch.f.}{characteristic function}
\acrodef{AWGN}{additive white Gaussian noise}
\acrodef{SNR}{signal-to-noise ratio}
\acrodef{LRT}{likelihood ratio test}
\acrodef{DRT}{distance ratio test}
\acrodef{GLRT}{generalized likelihood ratio test}
\acrodef{CRLB}{Cram\'{e}r-Rao lower bound}
\acrodef{CRB}{Cram\'{e}r-Rao bound}
\acrodef{ZZLB}{Ziv-Zakai lower bound}
\acrodef{ZZB}{Ziv-Zakai bound}
\acrodef{LOS}{line-of-sight}
\acrodef{ToF}{time-of-flight}
\acrodef{NLOS}{non-line-of-sight}
\acrodef{GDOP}{geometric dilution of precision}
\acrodef{GPS}{Global Positioning System}
\acrodef{FIM}{Fisher information matrix}
\acrodef{PEB}{position error bound}
\acrodef{SPEB}{squared position error bound}
\acrodef{TOA}{time-of-arrival}
\acrodef{TOF}{time-of-flight}
\acrodef{WSN}{wireless sensor network}
\acrodef{MAC}{medium access control}
\acrodef{RSS}{received signal strength}
\acrodef{WAF}{wall attenuation factor}
\acrodef{TDOA}{time difference-of-arrival}
\acrodef{RF}{radiofrequency}
\acrodef{RTT}{round-trip time}
\acrodef{AOA}{angle-of-arrival}
\acrodef{MF}{matched filter}
\acrodef{ED}{energy detector}
\acrodef{ML}{maximum likelihood}
\acrodef{MSE}{mean-square error}
\acrodef{RMSE}{root-mean-square error}
\acrodef{LEO}{localization error outage}
\acrodef{ppm}{part-per-million}
\acrodef{ACK}{acknowledge}
\acrodef{UWB}{Ultrawide bandwidth}
\acrodef{TNR}{threshold-to-noise ratio}
\acrodef{LS}{least squares}
\acrodef{IR-UWB}{impulse radio UWB}
\acrodef{FCC}{Federal Communications Commission}
\acrodef{TH}{time-hopping}
\acrodef{PPM}{pulse position modulation}
\acrodef{MUI}{multi-user interference}
\acrodef{PDP}{power delay profile}
\acrodef{BPZF}{band-pass zonal filter}
\acrodef{SIR}{signal-to-interference ratio}
\acrodef{SINR}{signal-to-interference-plus-noise ratio}
\acrodef{RFID}{radio frequency identification}
\acrodef{WPAN}{wireless personal area network}
\acrodef{WWB}{Weiss-Weinstein bound}
\acrodef{DP}{direct path}
\acrodef{MF}{matched filter}
\acrodef{MMSE}{minimum-mean-square-error}
\acrodef{SBS}{serial backward search}
\acrodef{SBSMC}{serial backward search for multiple clusters}
\acrodef{NBI}{narrowband interference}
\acrodef{WBI}{wideband interference}
\acrodef{INR}{interference-to-noise ratio}
\acrodef{CR}{channel response}
\acrodef{CIR}{channel impulse response}
\acrodef{CR}{channel  response}
\acrodef{RADAR}{radar}
\acrodef{MUR}{Multistatic radar}
\acrodef{JBSF}{jump back and search forward}
\acrodef{HDSA}{high-definition situation-aware}
\acrodef{RRC}{root raised cosine}
\acrodef{ST}{simple thresholding}
\acrodef{BTB}{Bellini-Tartara bound}
\acrodef{P-Max}{$P$-Max}  
\acrodef{MIMO}{multiple-input multiple-output}
\acrodef{MAP}{maximum a posteriori}
\acrodef{FG}{factor graph}
\acrodef{OP}{outage probability}
\acrodef{WED}{wall extra delay}
\acrodef{RMS}{root mean square}
\acrodef{SPAWN}{sum-product algorithm over a wireless network}
\acrodef{MDD}{minimum distance distribution}
\acrodef{MAP}{maximum a posteriori probability}
\acrodef{SAP}{small cell access point}
\acrodef{UE}{user equipment}
\acrodef{MBS}{macro cell base station}
\acrodef{UER}{\ac{UE} Relay}
\acrodef{D2D}{device-to-device}
\acrodef{MBS}{macro base station}
\acrodef{CSI}{channel state information}
\acrodef{OGR}{outage guard region}
\acrodef{FUR}{feasible UER region}
\acrodef{EHR}{energy harvesting region}
\acrodef{EH}{energy harvesting}
\acrodef{D2D-EHSN}{D2D communication provided \ac{EH} small cell network}
\acrodef{D2D-EHHN}{D2D communication provided \ac{EH} heterogeneous network}
\acrodef{3GPP}{3rd Generation Partnership Project}
\acrodef{BS}{base station}
\acrodef{DF}{decode and forward}
\acrodef{CCDF}{complementary cumulative distribution function}
\acrodef{ZF}{zero forcing}
\acrodef{RZF}{regularized zero forcing}
\acrodef{WLLN}{weak law of large number}
\acrodef{SLLN}{strong law of large numbers}
\acrodef{TDD}{Time-division duplex}
\acrodef{EE}{energy efficiency} 
\acrodef{HetNet}{heterogeneous network} 
\acrodef{SCP}{Single Cell Processing}
\acrodef{CBF}{Coordinated Beamforming}
\usepackage{color}
\usepackage{dsfont}
\usepackage{bbm}

\def\PST{P_{\mathrm{st}}}

\DeclareMathAlphabet{\mathsf}{OML}{cmbr}{m}{it}

\newtheorem{definition}{\bf Definition}
\newtheorem{theorem}{\bf Theorem}
\newtheorem{lemma}{\bf Lemma}
\newtheorem{corollary}{\bf Corollary}
\newtheorem{proposition}{\bf Proposition}

\newtheorem{remark}{\bf Remark}

\newtheorem{assumption}{\bf Assumption}

\newcommand{\bd}{\begin{description}}
\newcommand{\ed}{\end{description}}
\newcommand{\be}{\begin{enumerate}}
\newcommand{\ee}{\end{enumerate}}
\newcommand{\bi}{\begin{itemize}}
\newcommand{\ei}{\end{itemize}}
\newcommand{\bl}{\begin{list}}
\newcommand{\el}{\end{list}}
\newcommand{\bt}{\begin{tabbing}}
\newcommand{\et}{\end{tabbing}}

\newcommand{\paperTitle}{ A Unified Framework for SINR Analysis in Poisson Networks with Traffic Dynamics }

\begin{document}

{
\title{\paperTitle}

\author{

	    Howard~H.~Yang, \textit{Member, IEEE},
		Tony~Q.~S.~Quek, \textit{Fellow, IEEE},
		and H.~Vincent~Poor, \textit{Fellow, IEEE}



\thanks{

H.~H.~Yang and T.~Q.~S.~Quek are with the Information System Technology and Design Pillar, Singapore University of Technology and Design, Singapore (e-mail: howard\_yang@sutd.edu.sg, tonyquek@sutd.edu.sg).

H.~V.~Poor is with the Department of Electrical Engineering, Princeton University, Princeton, NJ 08544 USA (e-mail: poor@princeton.edu).
 }
}
\maketitle
\acresetall
\thispagestyle{empty}
\begin{abstract}
We study the performance of wireless links for a class of Poisson networks, in which packets arrive at the transmitters following Bernoulli processes.
By combining stochastic geometry with queueing theory, two fundamental measures are analyzed, namely the transmission success probability and the meta distribution of signal-to-interference-plus-noise ratio (SINR).
Different from the conventional approaches that assume independent active states across the nodes and use homogeneous point processes to model the locations of interferers, our analysis accounts for the interdependency amongst active states of the transmitters in space and arrives at a non-homogeneous point process for the modeling of interferers' positions, which leads to a more accurate characterization of the SINR.
The accuracy of the theoretical results is verified by simulations, and the developed framework is then used to devise design guidelines for the deployment strategies of wireless networks.
\end{abstract}

\acresetall
\section{Introduction}\label{sec:intro}
In recent years, there has been considerable progress toward understanding the performance of wireless links in large-scale networks by using tools from stochastic geometry \cite{HaeAndBac:09,Hae:12,BacBla:09,ElSSulAlo:17}.
By modeling the locations of transmitter-receiver pairs as spatial point processes, one can obtain simple expressions for a variety of key network statistics, e.g., coverage, throughput, or delay \cite{BacBla:09}, by capturing the spatial and physical layer attributes.
This intrinsic elegance has made stochastic geometry a disruptive method for performance evaluation among various wireless systems \cite{HaeAndBac:09,AndBacGan:11,DhiGanBac:12,YanLeeQue:16,KalHae:18}.
However, the majority of these stochastic geometry based analysis heavily relies on the \textit{full buffer} assumption, i.e., every link is active in the network, and do not allow one to represent temporal attributes such as packet generation and queue occupation.
Clearly, for complete network analysis, location tells just half the story and traffic assessment is of necessity.
To that end, the main purpose of this paper is to develop an analytical framework for the understanding of the impacts of spatial topology and temporal traffic dynamics, as well as their interdependence, on the link performance of a wireless network.

\subsection{Motivation and Related Work}
The main impediment of incorporating traffic dynamics into stochastic geometry based frameworks stems from the interdependency amongst the queueing evolutions, which is commonly known as the spatially interacting queues \cite{rao1988stability}.
Particularly, because wireless communications are conducted over a shared spectrum, transmissions in space will couple with each other via the interference they cause. In consequence, the evolution of queue at a given transmitter is fully entangled with those of its geographic neighbors, hence imposing a causality problem on the space-time interactions of the queues \cite{GhaElsBad:17}.
Understanding these causative interactions isn't easy, but it holds much of the key to understanding and coping with the design questions in wireless networks \cite{SanBac:17}.

In response, a recent line of studies has been conducted \cite{ZhoQueGe:16,ZhoHaeQue:16,GhaElsBad:17,yang2017packet,LiuZhoYan:17,LiHuaHan:18,GeoSpyKal:17,YanQue:19ICC,YanQue:19,ChiElSCon:17,ChiElSCon:19}, where stochastic geometry is combined with queueing theory to develop spatiotemporal models for large-scale wireless systems.
The particular approaches taken by these works can be classified into the following categories:
\begin{itemize}
\item[$a$)] \textit{Favorable/Dominant System Argument} \cite{ZhoQueGe:16,ZhoHaeQue:16}: This approach puts the focus on deriving bounds for the transmission success probability and delay. Specifically, by considering a favorable system where transmitters send out packets without retransmissions, an upper (resp. lower) bound can be derived for the transmission success probability (resp. delay). Analogously, by considering a dominant system where every transmitter is backlogged, lower (resp. upper) bounds are attainable for the transmission success probability (resp. delay). However, the favorable/dominant systems are often either too optimistic or too pessimistic compared to the real setup and hence result in, as we will show later loose bounds.
\item[$b$)] \textit{Stationary Approximation} \cite{yang2017packet,LiuZhoYan:17,LiHuaHan:18}: This approach evaluates the network performance under very light traffic condition, in which the majority of queues are stable. In this context, simple expressions are attainable for a number of network statistics, including the SINR coverage probability \cite{yang2017packet}, throughput \cite{LiuZhoYan:17}, and stable conditions \cite{LiHuaHan:18}.
    However, the accuracy of the analysis decays rapidly with an increase in the traffic load. Because that prolongs the active period of transmitters which in turns rise up the interference level thus incur many queues to switch from stationary into non-stationary regimes.
\item[$c$)] \textit{Geo/PH/1 Model} \cite{GhaElsBad:17}: This approach borrows advanced models from queueing theory and treats the geometry-dependent departure process as a phase type (PH) queue.
    The casuality of queueing interactions is then abstracted into a system of fixed point equations whose solutions can be used to evaluate the performance of different transmission schemes in large networks in terms of coverage and delay. The framework developed in \cite{GhaElsBad:17} is particular relevant to the uplink transmissions of narrow-band Internet-of-Things (NB-IoT), for which the single queue analysis is shown to attain good accuracy due to random codes.
    However, the analysis is carried out using the spatially averaged performance for all coexisting transmitters, which does not always result in an accurate estimation \cite{ChiElSCon:17,ChiElSCon:19}.
\item[$d$)] \textit{Meta Distribution Based Analysis} \cite{YanQue:19ICC,YanQue:19,ChiElSCon:17,ChiElSCon:19}: This approach utilizes the meta distribution of SINR to capture the diverse qualities of different transmission links and arrives at a refined characterization of the buffer-nonempty probability of each link. As a result, the coverage probability, as well as meta distribution of SINR, can be derived to quantify different levels of the quality-of-service (QoS) under various network models, ranging from cellular networks\cite{YanQue:19ICC,YanQue:19}, Poisson bipolar networks\cite{ChiElSCon:17}, to those with power controls and multiple channel access \cite{ChiElSCon:19}.
    However, as pointed out by \cite{ZhaYanShe:19}, the accuracy of these results deteriorates when the network is operating under a high SINR threshold or the infrastructure is densely deployed.
\end{itemize}
Whilst the details vary from one approach to another in the aforementioned categories, the analysis is commonly carried out utilizing double averaging \cite{Blas:2014user}: over time and network geometry. Specifically, the \textit{mean-field approximation}\cite{SioYua:12}, which assumes the queues evolve independently of each other, is first adopted to decouple the correlations in the packet dynamics at each node. Then, one can leverage the Little's law from classical queueing theory to calculate the mean active rates at individual link pairs, and, by conditioning on the positions of transceivers, obtain the time-averaged transmission success probability of each node.
Finally, the stochastic geometry is employed to average out the spatial randomness, and analytical expressions for the SINR related metrics can be subsequently obtained. In these procedures, although assuming the queues evolve independently over time is of necessity toward a tractable analysis, the time-averaged transmission success probabilities, which are often in the form of a system of fixed-point equations, remain mutually dependent in space.
However, the previous analysis implicitly assumes the distributions of these transmission success probabilities are independent and identical distributed (i.i.d.) across the transmitters, which result in homogeneous spatial distributions of interferers' locations.
In networks with sparsely deployed infrastructures, such approximation is justifiable because the mutual interference between any pair of transmitters is relatively ``weak'' and hence the interactions can be viewed as in a ``global'' manner.
However, as the network density increases, which is an inexorable trend of modern architecture \cite{LopDinCla:15}, transmitters in proximity will incur strong mutual interference and present a non-negligible correlation in their buffer status, making the interactions ``local''.
As such, adopting homogeneous models in the spatial averaging step, as already pointed out, lead to a potential repercussion of inaccurate analysis with which network designers bear the risk of making misleading conclusions.
Recognizing such constraint from the conventional tools, the central thrust of this paper is to improve the analysis of SINR from a joint queueing-geometry perspective -- by accounting for not just the spatial and temporal randomness, but also the interdependency amongst queue status -- such that the results can be used at any particular scale of a wireless network.

\subsection{Approach and Summary of Contributions}
In this paper, we deploy the transmitter-receiver dipoles as a Poisson bipolar network, in which the locations of transmitters follow a homogeneous Poisson point process (PPP) and each transmitter has a receiver at a fixed distance with random orientation\footnote{Note that such a setting is a large-scale analog to the classical model of \textit{Random Networks} \cite{GupKum:00}, in which the distance between any transmitter-receiver pair is fixed to represent the average value. Nevertheless, building upon the results from \cite{GhaElsBad:17} and \cite{YanQue:19}, the analysis developed in this paper can be extended to investigate networks with centralized infrastructures and multiple access/broadcast channels where transmitters are located at random distances to their receivers.}. From an engineering point of view, this network model is relevant to applications like Device-to-Device (D2D) communications, mobile crowdsourcing, and Internet-of-Things (IoT), which do not require a centralized infrastructure.
We employ a discrete time queueing system to model the temporal dynamic whereas the packet arrivals at each transmitter follow independent Bernoulli processes.
Every transmitter in this network maintains an infinite capacity buffer to store the incoming packets.
At each time slot, transmitters with non-empty buffers send out the packets from head of the line. Transmissions are successful only if the SINR received at the destination nodes exceed a predefined threshold, upon which the packet can be removed from the buffer.
Because of the shared spectrum, buffer state at each transmitter is correlated with others.
We thus combine the stochastic geometry and queueing theory to characterize the interference based interactions amongst the queues.
Specifically, on the macroscopic scale, we use stochastic geometry to account for the mutual interference among the transmitting nodes.
On the microscopic scale, we adopt queueing theory to account for the per-node buffer state.
In consequence, we extract a non-homogeneous PPP from the homogeneous setup to model the locations of interferers.
And based on that we derive accurate expressions for several key performance metrics.
Our main contributions are summarized below.
\begin{itemize}
\item We derive a tractable expression to characterize the transmission success probability by taking into account not only the randomness from packet arrival and network topology, but more importantly, the coupling effect of the queue states in space.
\item We derive an analytical expression for the meta distribution of SINR, which provides refined information of the fraction of wireless links that achieve SINR at any desired levels.
\item Using the mathematical framework, we obtain the optimal deployment densities that maximize the spatial throughput under different traffic conditions. Moreover, the performance fluctuation, as well as the 95\%-likely rate, of the wireless links are shown to be significantly affected by the traffic dynamics, and hence calls for new designs that jointly optimize the network performance with respect to the space-time attributes.
\end{itemize}


Compared to the existing results \cite{YanQue:19ICC,YanQue:19,ChiElSCon:17,ChiElSCon:19} that rely on the mean field approximation from both spatial and temporal perspectives to approach tractable analysis, this work successfully accounts for the effects of spatially queueing interactions in the analysis of SINR and hence advances the toolset for performance evaluation of large-scale networks with traffic dynamics. The developed theories provide a useful method for network operators to cope with various planning and optimization problems.

We organize the rest of this paper as follows. The configuration of the network is detailed in Section II. In Section III, we present the analysis of the transmission success probability, as well as the SINR meta distribution, in Poisson networks with traffic dynamics.
The accuracy of our analysis is verified by simulations in Section IV, along with design insights drawn from numerical examples.
Finally, several concluding remarks are made in Section V.

\begin{table}
\caption{Notation Summary
} \label{table:notation}
\begin{center}
\renewcommand{\arraystretch}{1.3}
\begin{tabular}{c  p{6.0cm} }
\hline
 {\bf Notation} & {\hspace{2.5cm}}{\bf Definition}
\\
\hline
$\tilde{\Phi}$; $\lambda$ & PPP modeling the spatial positions of transmitters; transmitter deployment density \\
$\hat{\Phi}$; $\lambda$ & PPP modeling the spatial positions of receivers; receiver deployment density \\
$\Phi$ & Superposition of the PPPs $\tilde{\Phi}$ and $\hat{\Phi}$, i.e., $\Phi = \tilde{\Phi} \cup \hat{\Phi}$ \\
$P_{\mathrm{tx}}$; $\alpha$ &  Transmit power; path loss exponent \\
$\delta$ &  An auxiliary notion defined as $\delta = 2/\alpha$ \\
$\xi$; $\xi_{\mathrm{c}}$ & Packet arrival rate; critical arrival rate \\
$\theta$ & SINR decoding threshold \\
$r$ & Distance between a typical transmitter-receiver pair \\
$\zeta_{j,t}$ & Indicator of active state at link $j$ during time slot $t$, which takes value 1 if the queue is nonempty and 0 otherwise \\
$a_j^\Phi$ & Queue nonempty probability at transmitter $j$, conditioned on the point process $\Phi$ \\
$\mu_{0, t}^\Phi$ & Transmission success probability of the typical link $0$ at time slot $t$, conditioned on the point process $\Phi$ \\
$p_{\mathrm{s}}$; $F_\theta(\cdot)$ &  Transmission success probability; SINR meta distribution \\
\hline
\end{tabular}
\end{center}\vspace{-0.63cm}
\end{table}%


\section{System Model}\label{sec:sysmod}
In this section, we introduce the network topology and propagation model, the packet arrival and transmission protocol, as well as the concept of spatially interacting queues. The main notations used throughout the paper are summarized in Table~I.

\subsection{Network Structure}
We consider an ad-hoc wireless network in which nodes are scattered according to a Poisson bipolar network in the Euclidean plane. The locations of transmitters follow a homogeneous Poisson point process (PPP) $\tilde{\Phi}$ of spatial density $\lambda$. Each transmitter $X_i \in \tilde{\Phi}$ has a dedicated receiver whose location $y_i$ is at distance $r$ in a random orientation. According to the displacement theorem \cite{BacBla:09}, the locations $\hat{\Phi} = \{y_i\}_{i=0}^\infty$ also form a homogeneous PPP with spatial density $\lambda$.
A realization of the network configuration is shown in Fig.~\ref{fig:SysMod_M1}.
In this network, all the transmitters transmit with unified power $P_{\mathrm{tx}}$\footnote{ We unify the transmit power to keep the analysis tractable, it shall be noted that the results from this paper can be extended to account for power control via a similar approach as in \cite{GhaElsBad:17}. }.
We assume the signal propagated between any two nodes is affected by the small-scale Rayleigh fading, which is independent and identical distributed (i.i.d.) across space and time, and the large-scale path loss that follows a power law.
Moreover, the received signal is also subject to white Gaussian thermal noise with variance $\sigma^2$.
For the sake of analytical tractability, we adopt a co-channel deployment on the network, i.e., all the nodes share the same spectrum for transmissions \cite{IEEE802-15-04}.

We model the evolution of the queues as a discrete time system. In particular, we segment the time axis into equal-duration slots where the time to transmit a single packet takes exactly one slot.
The packet arrivals to the transmitters form a collection of i.i.d. Bernoulli processes\footnote{The Bernoulli processes are essentially Poisson processes projected in a discrete time setting \cite{HlaKotSte:99}, which is a reasonable choice for the modeling of traffic attributes. Note that this model can be easily extended to represent more complicated traffic patterns by versions like Markov-Modulated Bernoulli Process (MMBP) \cite{MicLae:97,ChoHwaJun:98}.
} of rate $\xi$.
All the incoming packets are stored in a single-server queue with infinite capacity under the first-come-first-serve (FCFS) discipline.
At each time slot, every transmitter with a nonempty buffer sends out one packet from the head of the line. The transmission succeeds if the signal-to-interference-plus-noise ratio (SINR) at the intended receiver exceeds a predefined threshold.
Upon successful reception, the receiver feedbacks an ACK and the packet can be discarded at the sender side. Otherwise, the receiver sends a NACK message and the packet is retransmitted in the next time slot.
We assume the ACK/NACK transmissions are instantaneous and error-free, as commonly done in the literature.
Because the time scale of packet transmissions is much smaller than the dynamic of spatial positions, we assume the network topology is static, i.e., the locations of transmitters and receivers are generated once and remain unchanged in all the time slots.

Using the Slivnyak's theorem \cite{BacBla:09}, we can concentrate on a \textit{typical} receiver $y_0$ that is located at the origin with a tagged transmitter situated at $X_0$. Then, if a packet is sent out by the transmitter at the beginning of time slot $t$, the signal will propagate to the receiver at the end of the same slot with an SINR as
\begin{align} \label{equ:SINR_expr}
\gamma_{0,t} = \frac{P_{\mathrm{tx}} H_{00} r^{-\alpha} }{ \sum_{ j \neq 0 } P_{\mathrm{tx}} H_{j0} \zeta_{j,t} \Vert X_j - y_0 \Vert^{-\alpha} + \sigma^2 }
\end{align}
in which $H_{ji} \sim \exp(1)$ is the channel fading from transmitter $j$ to receiver $i$, $\alpha$ denotes the path loss exponent, $\zeta_{j,t} \in \{ 0, 1 \}$ indicates the buffer state of transmitter $j$ at time slot $t$ is empty (in this case, $\zeta_{j,t}=0$) or not (in this case, $\zeta_{j,t}=1$).
\begin{figure}[t!]
  \centering{}

    {\includegraphics[width=0.95\columnwidth]{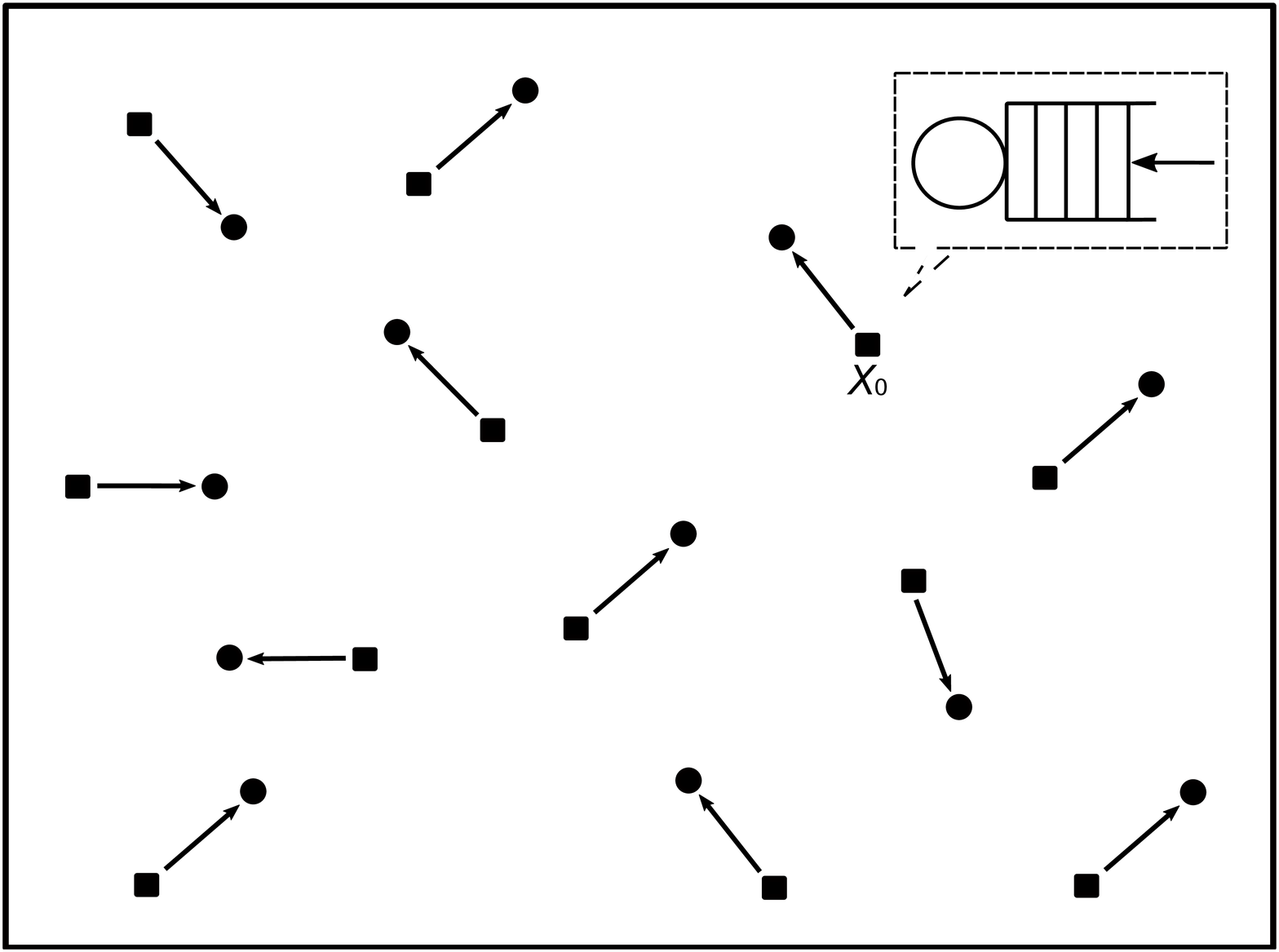}}

  \caption{ Example of a Poisson bipolar network with random traffic dynamics, the black squares and dots denote the transmitters and receivers, respectively. Each transmitter accumulates all the incoming packets into a buffer that has infinite size.
  }
  \label{fig:SysMod_M1}
\end{figure}

{\remark{\textit{Although this work is focused on dipoles, in similar spirits to \cite{BacBlaMuh:06,LiaRenLi:17}, one may extend the framework to consider multi-hop transmissions, given rise to a higher degree of spatial coupling amongst the queues. }}}

\subsection{ Spatially Interacting Queues }

In a wireless network, transmitters share the spectrum in space can impact each other's queueing states through the interference they cause.
As such, the active state of a generic link $j$, $\zeta_{j,t}$, is dependent on both the spatial and temporal factors.
A pictorial interpretation of this concept is given in Fig.~\ref{fig:QueItrc_M1}, which illustrates the spatiotemporal interactions among the queues of four wireless transmitter-receiver pairs.
From a spatial perspective, we can see that transmitters $X_1$ and $X_2$ are located in geographic proximity and hence their transmissions incur strong mutual interference, which slows down the rate of service\footnote{The service in this paper mainly refers to the packet transmission process, and hence the service rate is equivalent to the radiation rate which is only determined by the SINR.} and eventually prolongs their queue lengths.
In sharp contrast, transmitters $X_3$ and $X_4$ are at relatively long distances to their geographic neighbors.
Such advantageous locations benefit the transmissions in these links as they do not suffer severe crosstalk, and hence their buffer lengths are generally much shorter compared to those of transmitters $X_1$ and $X_2$.
From a temporal perspective, the packet arrival rate also plays a critical role in the process of service and further affects the queue length.
Particularly, if packets arrive at a high rate, all the transmitters will be active, which raises up the total interference level and that can incur many transmission failures, which prolong the active duration of the senders. On the contrary, when packet arrival rates are low, some transmitters may flush their queues and become silent, the reduced interference will also accelerate the depletion of packets at other nodes, which in turn leads to a shorten active period.

As such, in the context of a large-scale network, even if the packet arrivals are homogeneous in time, the spatial interactions result in a large variation of queue status across the nodes because the transmitters located in a crowded area of space will face poor transmission conditions and eventually have longer queue lengths than those situated at far distances from their neighbors.
Therefore, seen from any given link pair, the locations of interfering nodes are distributed inhomogeneously in space.
In the sequel, we aim to characterize this phenomenon into the analysis of SINR.

\begin{figure}[t!]
  \centering{}

    {\includegraphics[width=0.95\columnwidth]{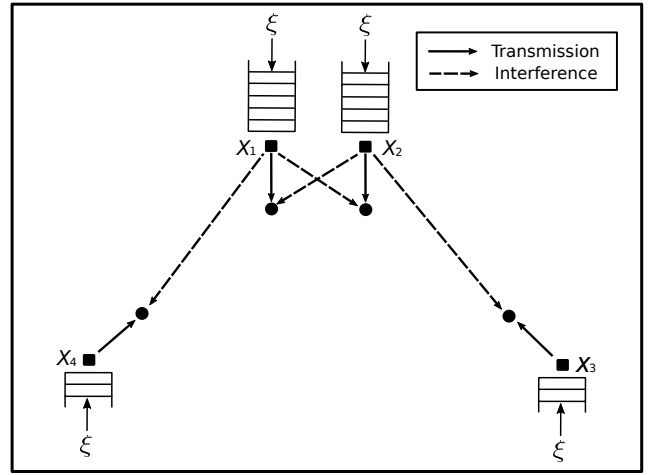}}

  \caption{ Illustration of a wireless network with spatially interacting queues. All the transmitter-receiver pairs are configured with the same distance and packet arrival rates. }
  \label{fig:QueItrc_M1}
\end{figure}

\subsection{Performance Metric}
In the rest of this paper, we elaborate the analysis on two fundamental metrics, i.e., the transmission success probability and the SINR meta distribution, that can be used to assess the network performance in terms of \textit{rate} and \textit{reliability}, respectively{\footnote{The main focus of this paper is on the SINR performance, although our model and its analysis can be carried out for any metrics regarding the delay in a straightforward way \cite{YanWanQue:18}.}}. More formal definitions are detailed below.
\subsubsection{Transmission success probability}In order to successfully deliver a packet within one time slot, the transmitters need to operate at certain rate level. Equivalently, that requires the SINR received at the destination nodes to exceed a decoding threshold. Because the SINR is governed by a number of random quantities, e.g., the channel fading and interference, we use the probability that the SINR is larger than a predefined threshold, usually referred to as the transmission success probability, to characterize this condition
\begin{align} \label{equ:Ps_AveCov}
p_{\mathrm s} = \mathbb{P}(\gamma_{0,t}>\theta).
\end{align}
This quantity can be thought of equivalently as, at any given time slot $t$, that
($i$) given an SINR target, the probability that a randomly chosen link can achieve successful transmission, or
($ii$) the average fraction of transmitters that can operate at SINR $\theta$.

\subsubsection{SINR meta distribution}Aside from the transmission success probability, statistics of the SINR can also be measured by a more fine-grained metric, namely the meta distribution \cite{haenggi2016meta}.
Formally, if we conditioned on the positions of the nodes $\Phi \triangleq \tilde{\Phi} \cup \hat{\Phi}$, the conditional transmission success probability of a typical link is given by
\begin{align}\label{equ: ConCovProb}
\mu^\Phi_{x_0, t} &= \mathbb{P}\left( \gamma_{x_0, t} \geq \theta | \Phi \right),
\end{align}
and the SINR meta distribution is defined as \cite{haenggi2016meta,ZhoHaeQue:16}:
\begin{align} \label{equ:F_meta_Genrl}
F_{\theta}(u) = \mathbb{P}\left( \mu^\Phi_{x_0, t} < u \right).
\end{align}
Different from \eqref{equ:Ps_AveCov} that gauges the average performance, \eqref{equ:F_meta_Genrl} allows one to obtain more subtle information such as the fraction of links that cannot achieve a certain transmission success probability and is often used for assessing the network reliability.

\section{ Analysis }
This is the main technical section of this paper, in which we derive analytical expressions for the transmission success probability as well as the SINR meta distribution in a general wireless network.

\subsection{Transmission Success Probability}
According to \eqref{equ:SINR_expr}, the SINR received at each link is dependent on the particular time slot as well as its relative location in the network, which can introduce memory in the queueing process via the spatiotemporal correlations and highly complicate the analysis.
That necessitates the introduction of the following approximation.
\begin{assumption}
\textit{
	In this network, each queue observes the time-averages of the activity indicators of other queues but evolves independently of their current state.
}
\end{assumption}

In essence, the approximation above makes the dynamic processes of packet transmissions conditionally independent, given positions of all transmitters and receivers, which is a mean-field approximation in the temporal domain. Consequently, we can put our focus on the asymptotic regime and drop the time index in the subsequential analysis.

In order to fully characterize the probability of successful transmissions, we first condition on the network topology $\Phi = \tilde{\Phi} \cup \hat{ \Phi }$ and average out the effect from the random channel fading.
When the network parameters are chosen to guarantee the stability of the queues, as will be detailed in Section III-B, a conditional form of the transmission success probability is attainable.
\begin{lemma}\label{lma:CondThrPut}
\textit{
	Given the spatial configuration $\Phi$, the probability of achieving successful transmissions over the typical link is given by
	\begin{align} \label{equ:CondThrPut}
	\mu_{0}^\Phi = e^{-\frac{ \theta r^\alpha }{\rho}}  \prod_{ j \neq 0 }  \bigg( 1 - \frac{ a_j^\Phi }{ 1 \!+\! \mathcal{D}_{j0}  } \bigg)
	\end{align}
	where $\rho = P_\mathrm{tx}/\sigma^2$, $\mathcal{D}_{ij} = \Vert X_i - y_j \Vert^\alpha / \theta r^\alpha$, and $a_j^\Phi = \lim_{ T \rightarrow \infty } \sum_{t=1}^{T} \zeta_{j,t}/T$ is the active probability of transmitter $j$ in the steady state.
}
\end{lemma}
\begin{IEEEproof}
Being conditional on the node positions, a packet delivered over the typical link can succeed with the following probability
\begin{align}
\mathbb{P} \left( \gamma_0 \!>\! \theta | \Phi  \right)
& =  \mathbb{P} \bigg(  { H_{00} }  >  \sum_{ j \neq 0 } \frac{ H_{j0} \zeta_{j,t}  \theta r^\alpha }{ \Vert X_j - y_0 \Vert^\alpha } + \frac{  \theta r^\alpha }{\rho} \,  \Big| \, \Phi  \bigg)
\nonumber\\
&= \mathbb{E}\bigg[ e^{-\frac{ \theta r^\alpha }{\rho}} \prod_{ j \neq 0 } \exp \Big( - \theta r^\alpha \frac{ H_{j0} \zeta_{j,t}  }{ \Vert X_j - y_0 \Vert^\alpha } \Big)  \Big| \Phi \bigg]
\nonumber\\
&\stackrel{(a)}{=} e^{-\frac{ \theta r^\alpha }{\rho}}  \prod_{j \neq 0} \Big( 1 - a_j^\Phi  + \frac{ a_j^\Phi }{ 1 + 1/\mathcal{D}_{j0} } \Big),
\end{align}
where ($a$) follows by using Assumption~1 and noticing that $H_{j0} \sim \exp(1)$.
The result can then be obtained by further simplifying the product factors.
\end{IEEEproof}

From \eqref{equ:CondThrPut}, we can immediately identify the randomness in the conditional SINR coverage probability, which mainly arises from ($a$) the random location of each transmitter and ($b$) its corresponding active state.
Furthermore, when conditioned on the point process $\Phi$, the packet transmission process at a generic link $j$ can be viewed as a Geo/Geo/1 queue with the rate of arrival and departure being $\xi$ and $\mu^\Phi_j$, respectively.
As such, by using the Little's law, we know the fraction of active period at link $j$ is given by
\begin{align} \label{equ:ActProb_Gnrl}
a^\Phi_j = \left \{
\begin{tabular}{cc}
\!\!\!\!\!\text{1}, & \text{if}~ $\mu^\Phi_j \leq  \xi$,   \\
\!\!\!\!  $\frac{ \xi }{ \mu^\Phi_j }$, & \text{if}~ $\mu^\Phi_j >   \xi$.
\end{tabular}
\right.
\end{align}

Putting \eqref{equ:CondThrPut} and \eqref{equ:ActProb_Gnrl} together, it is clear that the transmission success probability, as well as the active state, of any given node is a function of the transmission success probabilities of the others.
In other words, while Assumption~1 allows us to decouple the evolution of queues over time, the transmission success probability at individual links remains however mutually dependent in space due to interactions caused by interference.
Seen from the perspective of a typical transmitter, the level of mutual dependency to a given link $j$ can be reflected by the
active probability $a^\Phi_j$.
In particular, closer a link $j$ to the typical transmitter, higher their mutual interference and that leads to a larger value of $a^\Phi_j$, and vice versa.
Such interdependency between the transmitter active states and their geographic locations can be formalized in the lemma below.
\begin{lemma} \label{lma:DistCnd_ActProb}
\textit{When the typical link is activated\footnote{The reason of conditioning on the active state of the typical link is that the transmission success probability is evaluated as the number of successful transmissions over the total transmission times, which requires us to look at the transmission phase.}, given the transmission success probability $p_{ \mathrm{s} }$ and the distance between receivers $y_0$ and $y_j$ as $\Vert y_0 - y_j \Vert = u$, we have the following
	\begin{align} \label{equ:CondAct_j}
	&\mathbb{P}( \zeta_j = 1 \big| \Vert y_0 - y_j \Vert = u, \zeta_0 = 1 )
    \nonumber\\
	&\approx \int_0^{2\pi} \!\!\!\!\! \min \Big\{ \frac{ \xi }{ p_{ \mathrm{s} } } \big[ 1 + \frac{ \theta r^\alpha }{ ( u^2 + r^2 - 2 u r \cos \psi )^{ \frac{ \alpha }{ 2 } } }   \big], 1 \Big\} \frac{ d \psi }{ 2 \pi }.
	\end{align}
}
\end{lemma}
\begin{IEEEproof}
Please see Appendix~\ref{apx:DistCnd_ActProb}.
\end{IEEEproof}
\begin{figure}[t!]
  \centering{}

    {\includegraphics[width=0.95\columnwidth]{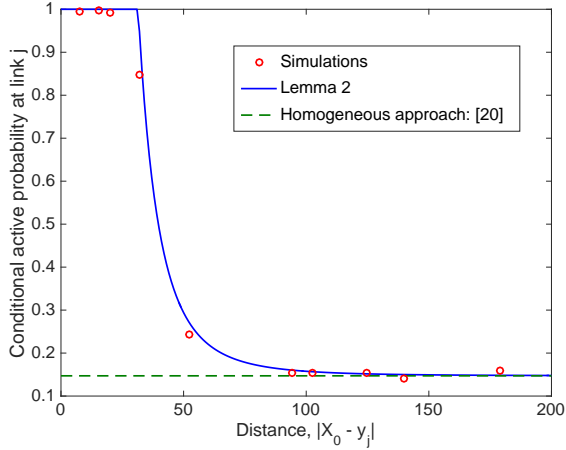}}

  \caption{ Conditional active probability at link $j$ vs the distance between receiver $j$ and the typical transmitter: $r=50$ m, $\xi=0.1$, $\alpha = 3.8$, and $\lambda = 10^{-4}~\mathrm{m}^{-2}$. }
  \label{fig:ActProb_Dist}
\end{figure}

The above result elucidates the change to the active probability at each link when the activation of a typical transmitter ripples through the network and deteriorates the transmissions of other nodes. In comparison to \cite{yang2017packet,YanQue:19,ChiElSCon:19} which assign universally equal active probability to each node, Lemma~\ref{lma:DistCnd_ActProb} quantifies the distance-dependent local interference in the active probability.
To better illustrate such differences, Fig.~\ref{fig:ActProb_Dist} plots the active probability at any given link $j$ as a function of the distance from receiver $j$ to the typical transmitter located at $X_0$, in which the simulations are drawn according to the setting in Section IV.
From this figure, we can see that the links in the vicinity of a typical transmitter have high probabilities of being activated, while the active states of links located far away are less affected.
In sharp contrast, the approaches in \cite{yang2017packet,YanQue:19,ChiElSCon:19} assume universally equal active probabilities at each transmitter, which underestimates the impact from interferers in the proximity and thus, as will be shown later, results in an upper bound of the SINR coverage probability.

We summarize the above discussions into the following proposition.
\begin{proposition}\label{prop:NonHomo_PPP}
\textit{
    Conditioned on the typical link being active, a transmitter-receiver pair located at $(X_j, y_j)$ activates with probability $G(u) = \mathbb{P}( \zeta_j = 1 \big| \Vert y_0 - y_j \Vert = u, \zeta_0 = 1 )$ and the propagation of interfering points constitute a non-homogeneous PPP with spatial density $\Lambda(u) = G(u) du$.
}
\end{proposition}

The resultant point process from Proposition~\ref{prop:NonHomo_PPP} has an intensity that decreases with increasing distance from the typical link. While the characterization of such an intensity is performed on a pair-wise basis, with focus on every single interfering link by averaging over the underlining Poisson configuration, it captures the -- what we term as -- ``first order interdependency'' and, as we will show later, leads to a good numerical result.
Using Proposition~\ref{prop:NonHomo_PPP}, we can conduct the computation via tools from non-homogeneous PPP, and that brings us to the main technical result of this paper.
\begin{theorem} \label{thm:ST_SINR_CoverProb}
\textit{
	The transmission success probability of the depicted wireless network can be approximated by the solution to the following fixed-point equation:
	\begin{align} \label{equ:CovProb_exct}
	p_{ \mathrm{s} } &\approx \exp\!\Big(\! -\! \frac{ \theta r^\alpha }{ \rho } \!-\! \lambda r^2 \!\!\! \int_0^\infty \!\!\!\! \int_0^{ 2\pi } \!\!\!\!\! \frac{ \mathcal{Z}(v, p_{\mathrm{s}}, \xi, \theta) v  d \varphi dv }{ 1 \!+\! ( 1 \!+\! v^2 \!-\! 2 v \cos \varphi )^{ \frac{ \alpha }{ 2 } }\!/\theta } \Big) \\ \label{equ:approx_CovProb}
	&\approx \exp\!\Big(\! -\! \frac{ \theta r^\alpha }{ \rho } \!-\! \lambda \pi r^2 \theta^\delta \!\!\! \int_0^\infty   \frac{ \min\!\big\{ \frac{ \xi }{ p_{ \mathrm{s} } } ( 1 \!+\! u^{ - \frac{ \alpha }{ 2 } } ), 1 \big\} }{ 1 + u^{ \frac{ \alpha }{ 2 } }  } du \Big)
	\end{align}
	where $\delta = 2 / \alpha$ and $\mathcal{Z}(v, p_{\mathrm{s}}, \xi, \theta)$ is given as follows:
	\begin{align}
	\mathcal{Z}(v, p_{\mathrm{s}}, \xi, \theta) =\!\! \int_0^{2\pi} \!\!\!\!\!\! \min\! \Big\{ \frac{ \xi }{ p_{ \mathrm{s} } } \big[ 1 \!+\! \frac{ \theta }{ ( 1  \!-\! 2 v \cos \psi \!+\! v^2 )^{ \frac{ \alpha }{ 2 } } }   \big], 1 \Big\} \frac{ d \psi }{ 2 \pi }.
	\end{align}
}
\end{theorem}
\begin{IEEEproof}
Please see Appendix~\ref{apx:ST_SINR_CoverProb}.
\end{IEEEproof}

Following Theorem~\ref{thm:ST_SINR_CoverProb}, a few observations can be immediately remarked.

\begin{remark}\textit{Due to the causative nature of spatially interacting queues, the transmission success probability is given in the form of a fixed-point (a.k.a. invariant point) equation.
}
\end{remark}

\begin{remark}\textit{The approximation in \eqref{equ:approx_CovProb} gives a lower bound to the transmission success probability and the approximation is tight when $\sqrt{\lambda} r \ll 1$.
}
\end{remark}


Note that the fixed-point equations \eqref{equ:CovProb_exct} and \eqref{equ:approx_CovProb} are numerically solvable for, and that several computing tools offer builtin routines, e.g., the fsolve function in Matlab, to accomplishing this task efficiently. Furthermore, \eqref{equ:CovProb_exct} also contains a special case version of closed-form solution:
\begin{corollary} \label{cor:SpCase}
\textit{When $\sqrt{\lambda} r \ll 1$, $\xi \ll 1$, and $\sigma^2 \ll \PST$, the transmission success probability can be approximated as follows:
\begin{align}
p_{ \mathrm s } \approx \frac{ - \lambda \xi \pi r^2 \theta^\delta \!\! \int_0^\infty \! \frac{ 1 + u^{ - \frac{\alpha}{2} } }{ 1 + u^{ \frac{\alpha}{2} } } \, du }{ \mathcal{W}\big( - \lambda \xi \pi r^2 \theta^\delta \!\! \int_0^\infty \! \frac{ 1 + u^{ - \frac{\alpha}{2} } }{ 1 + u^{ \frac{\alpha}{2} } } \, du \big) }
\end{align}
where $\mathcal{W}(\cdot)$ is the Lambert function \cite{AndAsk:00}.
}
\end{corollary}
\begin{IEEEproof}
On the one hand, when $\sqrt{\lambda} r \ll 1$ and $\sigma^2 \ll \PST$, the transmission success probability in \eqref{equ:CovProb_exct} can be tightly approximated as follows:
\begin{align} \label{equ:Aprx_v1}
p_{ \mathrm s } \approx \exp\!\Big( \!-\! \lambda \pi r^2 \theta^\delta \!\!\! \int_0^\infty \frac{ \min\big\{ \frac{ \xi }{ p_{ \mathrm s } } ( 1 + u^{ - \frac{\alpha}{2} } ), 1 \big\} }{ 1 + u^{ \frac{\alpha}{2} } } \Big).
\end{align}
On the other hand, when $\xi \ll 1$, the network can be approximated as stationary. Therefore, we have the following holds
\begin{align} \label{equ:Aprx_v2}
\min\big\{ \frac{ \xi }{ p_{ \mathrm s } } ( 1 + u^{ - \frac{\alpha}{2} } ), 1 \big\} \approx \frac{ \xi }{ p_{ \mathrm s } } ( 1 + u^{ - \frac{\alpha}{2} } ).
\end{align}
The result then follows from substituting \eqref{equ:Aprx_v2} into \eqref{equ:Aprx_v1} and perform further algebraic manipulation.
\end{IEEEproof}

Corollary~\ref{cor:SpCase} clearly shows the joint effect from spatial and temporal domains on the transmission success probability. Particularly, we note that under low traffic profile, the packet arrival rate $\xi$ and the deployment density $\lambda$ affect the probability of successful transmissions at the same level.

\subsection{ Stable Conditions }
From the temporal perspective, transmissions on each wireless link in the employed network can be abstracted as a queueing system in which the service rate is determined by the SINR statistics (i.e., the transmission success probability) at the intended receiver.
If the queues evolve in an isolated environment, then the stability can be guaranteed via the Loynes' theorem \cite{Loy:62}, by restricting the packet arrival rate to not exceed the average departure rate.
However, this condition cannot be directly extended to the depicted system where infinitely many transmitters interact with each other.
In fact, owing to the irregularity of the infrastructure, there are always some transmitters located in a congested spatial area with unbounded queue lengths. To that end, in lieu of restricting every individual queue to be stable, which can only be achieved under trivial circumstances (i.e., either $\xi = 0$ or $\theta = 0$), we opt for an alternative condition that keeps the fraction of unstable queues below an acceptable threshold. Formally, this is described by the concept of $\varepsilon$-stability \cite{ZhoHaeQue:16}.

\begin{definition}
\textit{For any $\varepsilon \in [0, 1]$, the $\varepsilon$-stable region $\mathcal{S}_{\varepsilon}$ is defined as
\begin{align} \label{equ:Gnrl_StableCond}
\mathcal{S}_{\varepsilon} = \bigg\{ \xi \in \mathbb{R}^+: \mathbb{P} \Big( \lim_{ T \rightarrow \infty } \frac{1}{T} \sum_{t=1}^{T} \mu^\Phi_{0,t} \leq \xi \Big) \leq \varepsilon \bigg\}
\end{align}
and the critical arrival rate $\xi_{ \mathrm{c} }$ is given as
\begin{align}
\xi_{ \mathrm{c} } = \sup \mathcal{S}_{ \varepsilon }.
\end{align}
}
\end{definition}
According to \eqref{equ:Gnrl_StableCond}, we know that when $\xi \leq \xi_{ \mathrm{c} }$, at most $\varepsilon$ fraction of the links are unstable. By setting $\varepsilon$ at a small value, it can then guarantee the majority of links have stable queues. Nonetheless, an exact expression of the critical arrival rate $\xi_{ \mathrm{c} }$ is still an open question, we thus resort to a few bounding techniques to find an approximation for the stable conditions.
\begin{theorem}
\textit{The sufficient condition for the network to remain $\varepsilon$-stable is
\begin{align} \label{equ:SufCnd}
\xi \leq \xi^{\mathrm{Sc}}_{\mathrm{c}} &= \sup\! \bigg\{ \xi \! \in \! \mathbb{R}^+\!\!: \frac{1}{2} -\!\! \int_{0}^{\infty} \!\!\!\!\!\! \mathrm{Im}\bigg\{ \xi^{ - j \omega } \exp \! \bigg(\!\! - \frac{j \omega \theta r^\alpha }{ \rho }
\nonumber\\
& \qquad \qquad -  \frac{ \lambda \pi^2 r^2 \delta^2 }{ \sin( \pi \delta ) }  \sum_{k=1}^{\infty} \! \binom{j \omega}{ k } \! \binom{ \delta \!-\! 1 }{ k \!-\! 1 } \! \bigg) \! \bigg\} \frac{ d \omega }{ \pi \omega } \leq \varepsilon \bigg\}
\end{align}
where $j = \sqrt{-1}$ and $\mathrm{Im}\{ \cdot \}$ denotes the imaginary part of a complex variable, and the necessary condition for the network to remain $\varepsilon$-stable is
\begin{align} \label{equ:NesCnd}
\xi \leq \xi^{\mathrm{Nc}}_{\mathrm{c}} &= \sup\! \bigg\{ \xi \! \in \! \mathbb{R}^+\!\!: \frac{1}{2} -\!\! \int_{0}^{\infty} \!\!\!\!\!\! \mathrm{Im}\bigg\{ \xi^{ - j \omega } \exp \! \bigg(\!\! - \frac{j \omega \theta r^\alpha }{ \rho }
\nonumber\\
&\qquad \quad -  \frac{ \lambda \pi^2 r^2 \delta^2 }{ \sin( \pi \delta ) }  \sum_{k=1}^{\infty} \! \xi^k \! \binom{j \omega}{ k } \! \binom{ \delta \!-\! 1 }{ k \!-\! 1 } \! \bigg) \! \bigg\} \frac{ d \omega }{ \pi \omega } \leq \varepsilon \bigg\}.
\end{align}
}
\end{theorem}
\begin{IEEEproof}
According to \eqref{equ:Gnrl_StableCond}, the condition of the network to remain $\varepsilon$-stable can be equivalently written as follows:
\begin{align} \label{equ:equiv_stbl_cndt}
\xi \leq \sup \mathcal{S}_{\varepsilon} = \sup \Big\{ \xi \in \mathbb{R}^+: \mathbb{P}\big( \mu^\Phi_0 \leq \xi \big) \leq \varepsilon \Big\}.
\end{align}
To obtain the sufficient condition, let us consider a dominant system, in which all the links are active regardless of the buffer states at the transmitters (if the buffer of a given node becomes empty, a ``dummy packet" will be sent out). Because transmissions in this system undergoes a higher level of interference than the original one, if $\varepsilon$-stability can be achieved in this system, it is also guaranteed under the original one.
Under the dominant system, each link is active and hence the conditional transmission success probability $\hat{\mu}^\Phi_0$ can be obtained by assigning $a^\Phi_j = 1, \forall j$ in \eqref{equ:CondThrPut}. As such, we can evaluate the $s$-th moment of $\hat{\mu}^\Phi_0$ as follows:
\begin{align} \label{equ:DmntSysm_smnt}
&\mathbb{E}\big[ (\hat{\mu}^\Phi_0)^s \big] 
\nonumber\\
&= e^{ - \frac{ s \theta r^\alpha }{ \rho } } \mathbb{E}\Big[ \prod_{ j \neq 0 } \big( 1 - \frac{1}{ 1 + \Vert X_j \Vert^\alpha/\theta r^\alpha } \big)^s \Big]
\nonumber\\
&= e^{ - \frac{ s \theta r^\alpha }{ \rho } } \exp\!\Big(\! - \lambda \int_{ \mathbf{x} \in \mathbb{R}^2 } \!\!\!\!\!\! \big[ 1 - \big( 1 - \frac{1}{ 1 + \Vert \mathbf{x} \Vert^\alpha/\theta r^\alpha } \big)^s \big] d \mathbf{x} \Big)
\nonumber\\
&= \exp\left( - \frac{ s \theta r^\alpha }{ \rho } - \lambda \pi \delta \frac{\pi \theta^\delta r^2 }{ \sin( \pi \delta ) } \sum_{k=1}^{s} \! \binom{ s }{ k } \! \binom{ \delta \!-\! 1 }{ k \!-\! 1 } \! \bigg)  \right).
\end{align}

By using the Gil-Pelaez theorem \cite{Gil}, we have
\begin{align} \label{equ:CDF_muPhi}
\mathbb{P}( \hat{\mu}^\Phi_0 < \xi ) = \frac{1}{2} - \frac{1}{\pi} \int_{0}^{\infty} \mathrm{Im}\Big\{ \xi^{ - j \omega } \mathbb{E}\big( \hat{\mu}^\Phi_0 \big)^{ j \omega } \Big\} \frac{ d \omega }{ \omega }.
\end{align}
The sufficient condition can then be obtained by substituting \eqref{equ:CDF_muPhi} and \eqref{equ:DmntSysm_smnt} back into \eqref{equ:equiv_stbl_cndt}.

Next, to obtain the necessary condition, we consider a favorable system where, at each node, every incoming packet is sent out once and discarded without retransmission. In this context, each link experiences a lower level of interference than that in the original system. Therefore, if the original system is $\varepsilon$-stable, the favorable system will follow suit.
And that constitutes the necessary condition. Note that the conditional transmission success probability, $\check{\mu}^\Phi_0$, of the favorable system can be obtained by taking $a^\Phi_j = \xi, \forall j$ in \eqref{equ:CondThrPut} and the derivation of $\varepsilon$-stability follows a similar approach as above.
\end{IEEEproof}

The critical arrival rates $\xi^{ \mathrm{Sc} }_{ \mathrm{c} }$ and $\xi^{ \mathrm{Nc} }_{ \mathrm{c} }$ in \eqref{equ:SufCnd} and \eqref{equ:NesCnd}, respectively, defines the boundaries in which the largest possible arrival rate, upon which the network remains stable, lies in. To calculate these quantities, we need to solve for inequalities where $\xi$ appears at the both sides.
 \newcounter{TempEqCnt}
  \setcounter{equation}{\value{equation}}
  \setcounter{equation}{21}
  \begin{figure*}[t!]
  \begin{align} \label{equ:Meta_Grl}
  F_{\theta}(u) &= \frac{1}{2} -\! \int_{0}^{\infty} \!\!\! \mathrm{Im}\bigg\{ u^{-j\omega} \exp\!\Big(\! - \frac{ j \omega \theta r^\alpha }{ \rho } - 2 \lambda \pi r^2 \sum_{k=1}^{\infty} \binom{j \omega}{ k } (-1)^{k+1} \!\!\int_0^\infty \!\!\! \int_0^{2\pi} \!\! \big[ 1 - \frac{ \xi }{ \mathcal{H}_\theta(v,\psi) } \big]^k \frac{d \psi}{2\pi}
  \nonumber\\
 &\qquad \qquad \qquad \qquad \qquad \qquad \qquad \quad \times \int_{0}^{2\pi} \Big[ F_{\theta}\big( \mathcal{H}_\theta(v,\varphi) \big) \!+\!\! \int_{ \mathcal{H}_\theta(v,\varphi) }^{1}  \!\!\!\!\!\!  \frac{ \mathcal{H}^k_\theta(v,\varphi) }{t^k} \, F_{ \theta }(dt)
  \Big] \frac{ d \varphi }{ 2\pi }  v dv \Big) \bigg\} \frac{ d \omega }{ \pi \omega }
  \end{align}
  \setcounter{equation}{\value{equation}}{}
  \setcounter{equation}{22}
  \centering \rule[0pt]{18cm}{0.3pt}
  \end{figure*}
  \setcounter{equation}{22}

\subsection{ SINR Meta Distribution }
We now turn our attention to the aspect of network reliability and derive the expression for the meta distribution of SINR.
It is worth noting that compared to the transmission success probability, which provides information about the average, the SINR meta distribution answers more fine-grained questions, for instance: ``How are the transmission success probabilities of individual links distributed in a realization of the Poisson network?" which directly leads to the performance of, e.g., the top 95\% of transmitters, and is an important design criterion for network operators.
\begin{theorem} \label{thm:Meta_SINR}
  \textit{The SINR meta distribution is given by the fixed-point equation \eqref{equ:Meta_Grl} at the top of next page, 
  in which the auxiliary function $\mathcal{H}_\theta(x, y)$ is given as follows:
  \begin{align}\label{equ:H_theta}
    \mathcal{H}_\theta(x,y) = \xi + \frac{ \xi \theta  }{ ( 1 - 2 x \cos y + x^2 )^{ \frac{\alpha}{2} } }.
  \end{align}
  Furthermore, \eqref{equ:Meta_Grl} can be iteratively solved as follows:
  \begin{align}
  F_\theta(u) = \lim_{ n \rightarrow \infty } F_{ \theta, n }(u)
  \end{align}
  where $F_{ \theta, n }(u)$ is given by
  \begin{align} \label{equ:F_theta_n}
  F_{ \theta, n }(u) &= \frac{1}{2} - \! \int_0^\infty \!\!\!\!\! \mathrm{Im} \bigg\{ u^{-j\omega} \exp\!\Big(\! - \frac{ j \omega \theta r^\alpha }{ \rho } 
  \nonumber\\
  &\qquad - \frac{ \lambda r^2 }{ 2 \pi } \sum_{k=1}^{\infty} \binom{j \omega}{ k } (-1)^{k+1} \tilde{\eta}^{(k)}_{n-1} \Big) \bigg\} \frac{ d \omega }{ \pi \omega }
  \end{align}
  whereas $\tilde{\eta}^{(k)}_{n-1}$ is given by
  \begin{align}
  \tilde{\eta}^{(k)}_{n-1} & =\!\! \int_0^\infty \!\!\! \int_0^{ 2 \pi } \!\!\!\! \big[\, 1 - \frac{ \xi }{ \mathcal{H}_\theta( v, \psi ) } \, \big]^k d \psi \int_{0}^{2\pi}  \!\!\!\! \big[\, F_{ \theta, n-1 }\big( \mathcal{H}_\theta( v, \varphi ) \big)
\nonumber\\
& \qquad \qquad \qquad + \int_{ \mathcal{H}_\theta( v, \varphi ) }^{ 1 } \!\!\!\!  \frac{ \mathcal{H}^k_\theta( v, \varphi ) }{t^k} F_{ \theta, n-1 }( dt ) \big] d \varphi v dv.
  \end{align}
  If $n=1$, we have $\tilde{\eta}_0^{(k)}$ given as follows:
  \begin{align}
  \tilde{\eta}_0^{(k)} = \binom{ \delta - 1 }{ k - 1 } \frac{ 2 \pi^2 \delta \theta^\delta \xi^k }{ \sin( \pi \delta ) }
  \end{align}
}
\end{theorem}
\begin{IEEEproof}
Please see Appendix~\ref{apx:Meta_SINR}.
\end{IEEEproof}

Different from the analysis presented in \cite{YanQue:19}, the result in \eqref{equ:Meta_Grl} successfully captures the spatial interdependency of queue active states between a typical transmitter and its geographical neighbors and provides an expression for the SINR meta distribution computed from a \textit{non-homogeneous} PPP.
If we treat the dynamics on a typical link as a Geo/G/1 queue, function \eqref{equ:Meta_Grl} corresponds to the distribution of the service rate. And we can use it to assess the performance of time-domain metrics such as delay or throughput, though that is beyond the scope of this paper and leave as future works.
To carry out the computation, we can set an accuracy threshold $\epsilon$, which is sufficiently small, and stop the iteration when $\vert \tilde{\eta}_n^{(k)} - \tilde{\eta}_{n-1}^{(k)} \vert < \epsilon, \forall k$ where $\tilde{\eta}_n^{(k)}$ is given in (26). Actually, such a iteration can converge in very few, e.g., less than 10, steps as demonstrated in \cite{ZhaYanShe:19}.

While $F_\theta(u)$ can be solved in a recursive manner, each iteration requires the computation of all moments of the conditional transmission success probability which can be time consuming.
One way to get around this difficulty is to approximate the function $F_\theta(u)$ by a Beta distribution.
The detailed approaches are summarized in the following corollary:
\begin{corollary}\label{Cor:Approximation}
\textit{The probability density function (pdf) of $F_{\theta}(u)$ in Theorem~\ref{thm:Meta_SINR} can be tightly approximated via the following
\begin{align}
f_X(u) &= \lim_{ n \rightarrow \infty } f_{X_n}(u)
\nonumber\\
&= \lim_{ n \rightarrow \infty } \frac{u^{\frac{\mu_n (\beta_n + 1) - 1}{1 - \mu_n}} (1-u)^{\beta_n - 1} }{B(\mu_n \beta_n/(1-\mu_n), \beta_n )}
\end{align}
where $B(a, b)$ denotes the Beta function \cite{AndAsk:00}, $\mu_n$ and $\beta_n$ are respectively given as
\begin{align} \label{equ:mu_n}
\mu_n &= M_n^{(1)}, \\ \label{equ:beta_n}
\beta_n &= \frac{(\mu_n - M_n^{(2)}) ( 1 - \mu_n )}{M_n^{(1)} - \mu_n^2 }
\end{align}
where $M_n^{(m)}$ can be written as
\begin{align} \label{equ:Momnt_Beta}
& M_n^{(m)} \!=\! \exp\!\bigg(\!\! -\! \frac{ m \theta r^\alpha }{ \rho }  \! - \! \lambda r^2 \! \sum_{k=1}^{m} \! \binom{ m }{ k } (-1)^{ k+1 } \, \hat{\eta}^{(k)}_n \!\bigg),
\end{align}
and $\hat{\eta}^{(k)}_n$ is given by
\begin{align}
\hat{\eta}^{(k)}_{n-1} & =\!\! \int_0^\infty \!\!\! \int_0^{ 2 \pi } \!\!\!\! \big[\, 1 - \frac{ \xi }{ \mathcal{H}_\theta( v, \psi ) } \, \big]^k d \psi \!\! \int_{0}^{2\pi}  \!\!\!\! \big[\! \int_{0}^{ \mathcal{H}_\theta( v, \varphi ) } \!\!\!\!\!\!\!\!\!\!\!\!\!\!\!\!   f_{ X_{n - 1} }(t)dt 
\nonumber\\
&\qquad \qquad + \!\! \int_{ \mathcal{H}_\theta( v, \varphi ) }^{ 1 } \!\!\!\!\!   \frac{ \mathcal{H}^k_\theta( v, \varphi ) }{t^k} f_{ X_{n - 1} }(t)  dt \big] \frac{d \varphi}{ 2 \pi } v dv.
\end{align}
Particularly, when $n=1$, we have $\hat{\eta}_{0}^{(k)}$ given by the following
\begin{align}
\hat{\eta}_{0}^{(k)} = \binom{ \, \delta - 1 \,}{\, k - 1 \,} \frac{ 2 \pi^2 \theta^{ \delta } \xi^k  }{ \alpha \sin(  \pi \delta ) }.
\end{align}
}
\end{corollary}
\begin{IEEEproof}
It can be observed from \eqref{equ:F_meta_Genrl} that the approximated function $F_{\theta,n}(u)$ in each iteration step is supported on $[0,1]$. We are thus motivated to approximate the distribution via a Beta distribution.
First, using results in \eqref{equ:MmnGen_Y} we can derive the moments in \eqref{equ:Momnt_Beta}. Next, by respectively matching the mean and variance to a Beta distribution $B(a_n, b_n)$, it yields
\begin{align}
& \frac{ a_n }{ a_n + b_n } = M_n^{(1)}, \\
& \frac{ a_n b_n }{ ( a_n + b_n )^2 ( a_n + b_n + 1 ) } = M_n^{(2)} - \big[ M_n^{(1)} \big]^2
\end{align}
and the result follows from solving the above system equations.
\end{IEEEproof}

\section{Simulation and Numerical Results}
In this section, we validate the accuracy of our analysis through simulations and evaluate different network statistics based on the numerical results.
Particularly, we consider a square region with side length of 1 km, in which link pairs are scattered according to a Poisson bipolar network with spatial density $\lambda$ and once the topology is generated it remains unchanged.
To eliminate the favorable interference coordinations induced by network edges, we use wrapped-around boundaries \cite{FasMueRup:19} that allow dipoles that leave the region on one side to reappear on the opposite side, thus mirroring the missing interferers beyond the scenario boundary.
Then, the packet dynamics at each link are run over 10,000 time slots.
Specifically, at the beginning of each time slots, channel gains are independently instantiated and packets are generated at each sender with probability $\xi$. The nodes with non-empty buffers then send out packets according to a FCFS discipline with failure retransmission occur at the next time slot. And a packet can be dropped from the transmitter queue if the received SINR at the intended node exceeds the decoding threshold.
The SINR statistics of the receivers of all active links are recorded to construct the transmission success probability (calculated as the ratio between the number of successful transmissions over the total transmission times) as well as the meta distribution of SINR.
Unless otherwise stated, we set the system parameters as follows: $\alpha = 3.8$, $\theta=0$~dB,  $\xi=0.1$~packet/slot, $P_{\mathrm{tx}}=17$~dBm, $\sigma^2 = -90$~dBm, $r=25$~m, and $\lambda = 10^{-4}~\mathrm{m}^{-2}$.
\subsection{SINR and Rate Performance}
\begin{figure}[t!]
  \centering{}

    {\includegraphics[width=0.95\columnwidth]{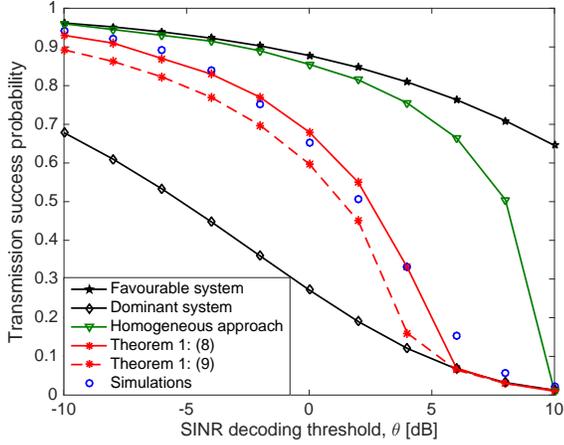}}

  \caption{ Transmission success probability vs detection threshold. }
  \label{fig:SINR_M1}
\end{figure}

In Fig.~\ref{fig:SINR_M1}, the simulated transmission success probability is compared to the analytical ones derived via different approaches.
Particularly, the analytical results are calculated by means of the
$a$) favorable/dominant system arguments \cite{ZhoQueGe:16},
$b$) homogeneous approach \cite{GhaElsBad:17,YanQue:19},
and $c$) analysis developed in Theorem~\ref{thm:ST_SINR_CoverProb}.
The figure shows that analytical results and simulations well match, validating the accuracy of Theorem~\ref{thm:ST_SINR_CoverProb}.
We also find that the upper and lower bounds derived according to \cite{ZhoQueGe:16} are not just loose, but more crucially, the tendency of the analysis deviates a lot from the simulations.
The reason is attributed to the fact that both the favorable and dominant systems are essentially modified versions of the conventionally full buffer assumption, and hence not capturing the intrinsic effect from the space-time interactions between the queues.
In fact, compared to the dominant system, the lower bound given in \eqref{equ:approx_CovProb} is much tighter because it takes into account the non-homogeneity property from the interference point process.
Similarly, we can see that the success probability derived under the homogeneous approach \cite{GhaElsBad:17,YanQue:19} also fails to characterize the true distribution.
This is because the spatial interactions of queues lead to a location-dependent active probability at each node, as illustrated per Fig.~\ref{fig:ActProb_Dist}, but the homogeneous approach assumes the active states are i.i.d. across transmitters and thus results in an underestimation of the interference.
In summary, through the space-time interactions, the resulting point process of the active nodes is non-homogeneous, and we can include this fact in the analysis to attain a comprehensive understanding of the transmission success probability.

\begin{figure}[t!]
  \centering{}

    {\includegraphics[width=0.95\columnwidth]{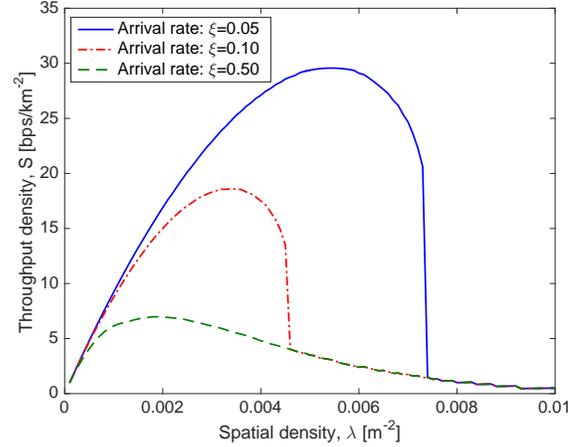}}

  \caption{ Throughput density vs network spatial density. }
  \label{fig:ActDen_M1}
\end{figure}

\begin{figure}[t!]
  \centering{}

    {\includegraphics[width=0.95\columnwidth]{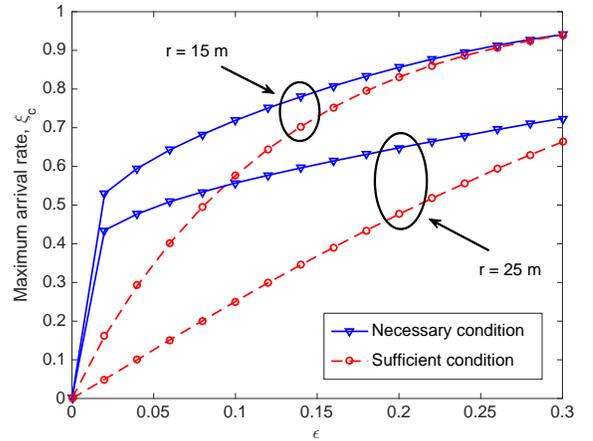}}

  \caption{ Comparison of sufficient and necessary conditions for $\varepsilon$-stability. }
  \label{fig:Stable_Region}
\end{figure}

Fig.~\ref{fig:ActDen_M1} depicts the throughput density \cite{BacBla:09}, defined as $\mathcal{S} = \lambda \cdot \log_2(1+\theta) \cdot \mathbb{P}( \gamma_0 > \theta )$, as a function of the spatial density.
From this figure, we can clearly observe the impacts of network parameters on the throughput density from the perspective of both space and time.
On the one hand, an optimal deployment density exists due to a tradeoff between the increasing number of active links and the rising interference power.
On the other hand, the traffic pattern also plays a critical role in determining the maximally achievable throughput density.
Specifically, in the light traffic regime, one can deploy a large number of transceiver pairs and attain high throughput density thanks to the low activity rate of transmitters.
In contrast, when the nodes are heavily loaded, both the optimal deployment density and throughput density drop quickly since most of the links are activated and the interference level is high.

\subsection{Stability and Reliability}
\begin{figure*}[t!]
  \centering

  \subfigure[\label{fig:1a}]{\includegraphics[width=0.95\columnwidth]{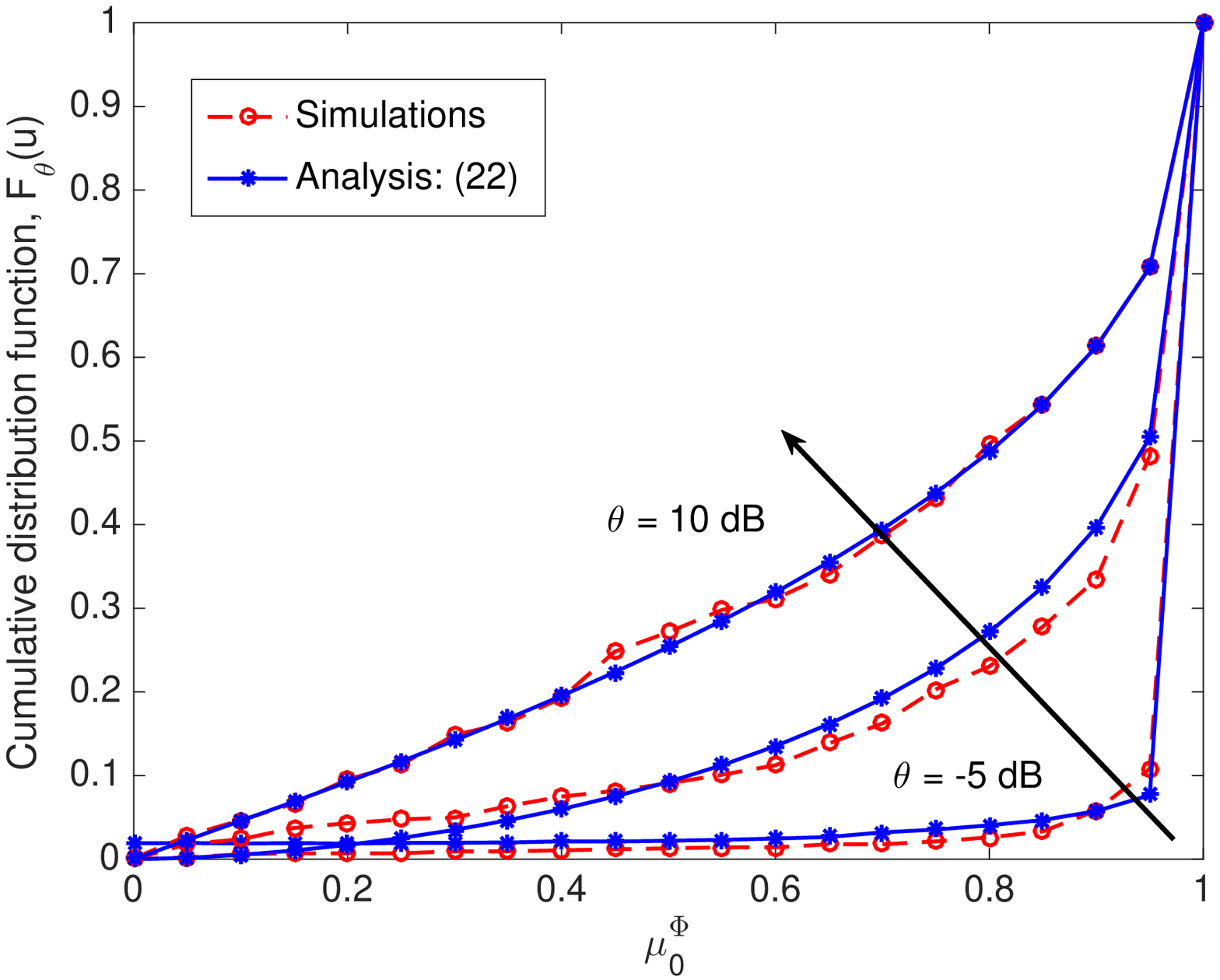}} ~
  \subfigure[\label{fig:1b}]{\includegraphics[width=0.95\columnwidth]{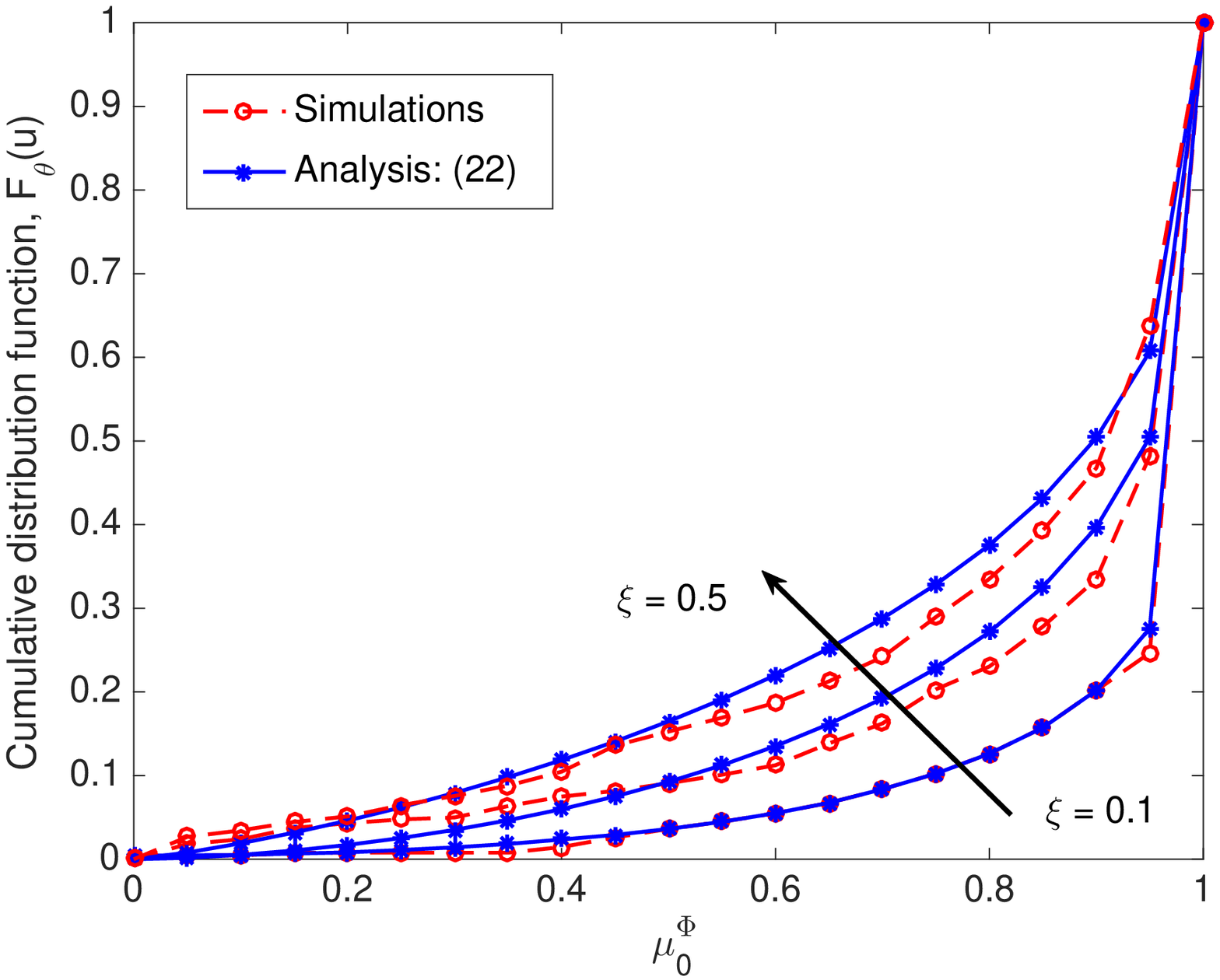}}
  \caption{Simulation versus analysis: SINR meta distribution. In Fig. (a), the packet arrival rate is fixed as $\xi = 0.1$, and the decoding threshold varies as $\theta = -5, 0, 10$~dB. In Fig. (b), the SINR threshold is set as $\theta = 0$~dB, while the packet arrival rates change according to $\xi = 0.1, 0.3, 0.5$. }
  \label{fig:meta_SINR}
\end{figure*}

Fig.~\ref{fig:Stable_Region} plots the critical arrival rates of both the sufficient and necessary conditions as functions of $\varepsilon$, under two sets of transmitter-receiver distance $r$, in which the critical arrival rate of the employed system lies in between.
The figure reveals that increasing the rate of packet arrival leads to a larger portion of queues being unstable. Note that if $\varepsilon$ is set to be 0.3, the maximum arrival rate can be high, i.e., $\xi_{\mathrm{c}} \approx 0.9$ for $r=15$~m and $\xi_{\mathrm{c}} \approx 0.7$ for $r=25$~m. This is because, in both scenarios, the distances between transmitter-receiver pairs are smaller than the average inter-link distance, which results in relatively high signal power at the receiver side. Nonetheless, it is more desirable to set the packet arrival rates at small so as to maintain the majority of all the queues stable in the network.

In Fig.~\ref{fig:meta_SINR}, we put the spotlight on the SINR meta distribution, with varying values of decoding threshold $\theta$ and packet arrival rate $\xi$. First of all, we note that the results obtained via simulations match well with those from Theorem~\ref{thm:Meta_SINR}, thus confirming the analysis. Next, we can use Fig.~\ref{fig:1a} to assess the confidential level about the network reliability under different rate thresholds. Specifically, for a decoding threshold of -5~dB, 93\% of the links can successfully achieve the targeted SINR with a probability of at least 0.90. However, when the decoding threshold raises to 10~dB, only 30\% of the links are able to attain successful transmissions with the same probability (i.e., 0.90).
This can serve as guidance for the operators to adjust the transmission rate targets in accordance with different levels of reliability.
Furthermore, results from Fig.~\ref{fig:1b} also shows the impact of temporal factors on the SINR. In particular, with an increase of packet arrival rate, the SINR will be deteriorated, which is reflected by a steady, but non-linear, uptrend to the meta distribution.
It can be seen that the change of SINR meta distribution is more noticeable when the packet arrival rate grows from small ($\xi=0.1$) to a medium value ($\xi=0.3$), and the trend slows down as the traffic load further increases (to $\xi=0.5$). This is because, on the one hand, many links are deactivated in the light traffic condition, as the packet arrival rate increases, it not just activates more transmitters, but more crucially, gives rise to a higher interference level. And that incurs more delivery failures and retransmissions which prolong the active duration of the nodes.
This composite effect accelerates the degradation of SINR across the nodes, leading to a sharp change of the meta distribution.
On the other hand, when the traffic load is relatively high, most of the queues are non-empty, the additional interference then contributes less to the total level, and thus the trend slows down.

\begin{figure}[t!]
  \centering{}

    {\includegraphics[width=0.95\columnwidth]{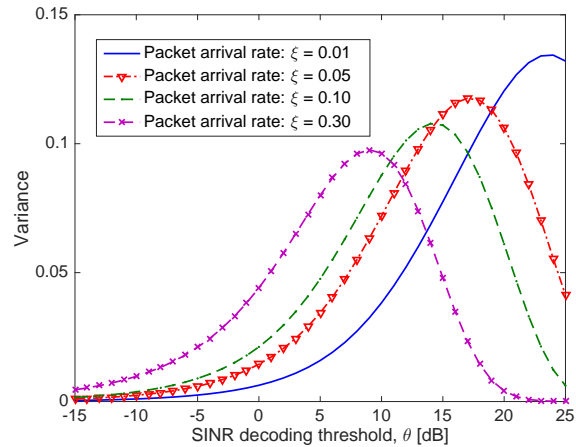}}

  \caption{ Variance of transmission success probability versus decoding threshold. }
  \label{fig:Var_v1}
\end{figure}

Fig.~\ref{fig:Var_v1} shows the variance of the transmission success probability as a function of $\theta$. Similar to [33], the value of variance in this figure is calculated by using the meta distribution of SINR given in (22).
Note that as the variance neccisarily tends to zero at the two extreme ends of $\theta$, i.e., $\theta \rightarrow 0$ or $\theta \rightarrow \infty$, it assumes the maximum at some finite value of $\theta$. Particularly, from this figure we can see that the network peaks at different variance as the traffic condition varies, therefore the performance fluctuation of the wireless links is directly affected by the traffic pattern.

\begin{figure}[t!]
  \centering{}

    {\includegraphics[width=0.95\columnwidth]{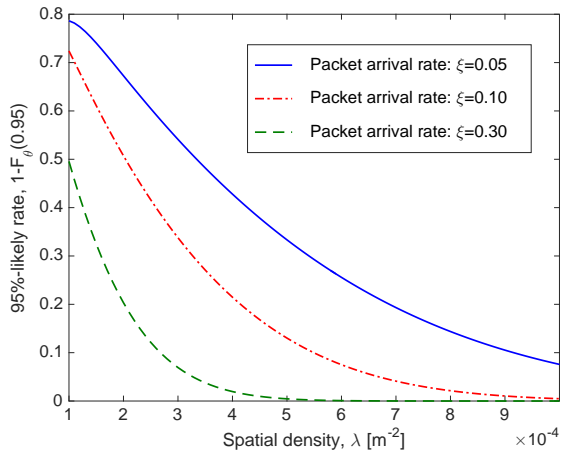}}

  \caption{ 95\%-likely rate versus spatial density. }
  \label{fig:96LR_v1}
\end{figure}
In Fig.~\ref{fig:96LR_v1} we plot the 95\%-likely rate, i.e., $1 - F_\theta(0.95)$, as a function of the spatial density $\lambda$.
This quantity gives information about the performance of the ``worst 5\% transmitters'', namely the link pairs in the bottom 5th percentile in terms of data rate performance, and is particularly interested to operators \cite{AndBuzCho:14,haenggi2016meta}.
We observe from Fig.~\ref{fig:96LR_v1} that $a$) the 95\%-likely rate declines precipitously when the network grows in size and $b$) an increase of the packet arrival significantly defects the 95\%-likely rate.
These observations confirm the intuition that the worst 5\% transmitters are also the most vulnerable ones to the change of space-time situation of a wireless network.
As such, developing advanced channel access mechanism to boost up the 95\%-likely rate is of necessity to enhance the overall network performance.

\section{Conclusion}
In this paper, we have introduced a mathematical toolset that allows one to evaluate the SINR performance of wireless networks from a space-time perspective.
Our model is general and accounts for a congeries of key features including the channel fading, path loss, network topology, traffic dynamics, and spatially interacting queues.
By jointly using queueing theory and stochastic geometry, we have characterized the locations of the interfering nodes to be a non-homogeneous PPP and obtained accurate expressions for both the transmission success probability and SINR meta distribution.
Based on the analysis, we obtained an optimal deployment density that achieved the maximum throughput density under different traffic conditions.
We also confirmed that the traffic pattern directly affects the performance fluctuation of wireless links.
Moreover, the analysis revealed that the worst 5\% transmitters are vulnerable to a change in space-time condition, and that calls for advanced technologies to accommodate the transmissions of these nodes.

The spatiotemporal framework established in this paper can facilitate the design and understanding of various wireless systems.
For stance, one can use it to devise channel access schemes for internet-of-things (IoT) networks with a guarantee to the latency and reliability or obtain accurate evaluation to the performance of next-generation wireless local access networks by taking into account the space-time queueing interactions.
Investigating to what degree the power controls affect the spatiotemporal analysis is also a concrete direction for future research.

\begin{appendix}
\subsection{Proof of Lemma~\ref{lma:DistCnd_ActProb} } \label{apx:DistCnd_ActProb}
Conditioned on the event that the typical link is active, i.e., $\zeta_0 = 1$, we can adopt Lemma~\ref{lma:CondThrPut} and rewrite the conditional SINR coverage probability at a given link $j$ in the following way
\begin{align}
\mu^\Phi_j &= e^{ -\frac{ \theta r^\alpha }{ \rho } } \! \prod_{ \substack{ i \neq 0, \\ i \neq j } } \! \big( \, 1 \!-\! \frac{ a_i^\Phi }{ 1 \!+\! \mathcal{D}_{ ij } } \, \big) \big( 1 \!-\! \frac{ 1 }{ 1 \!+\! \mathcal{D}_{0j} } \big)
\nonumber\\
&= \frac{ \mathcal{D}_{0j} }{ 1 + \mathcal{D}_{0j} } \times \mu^{\Phi^{!0}}_j,
\end{align}
where $\mu^{\Phi^{!0}}_j$ denotes the conditional transmission success probability of link $j$ given the point process $\Phi$ except the node at $X_0$, i.e., a reduced point process $\Phi^{!0}$ \cite{BacBla:09}.
Using the expression in \eqref{equ:ActProb_Gnrl}, we can then take an expectation with respect to $\Phi^{!0}$ in the conditional success probability and obtain the conditional active probability at link $j$ as follows:
\begin{align} \label{equ:Cnd_ActProb}
& \mathbb{P}( \zeta_j = 1 | \Vert y_0 - y_j \Vert = u, \zeta_0 = 1 )
\nonumber\\
& =  \mathbb{E}\Big[ \min\big\{ \big( 1 + \frac{ 1 }{ \mathcal{D}_{0j} } \big) \cdot \mathbb{E}\big[ \frac{ \xi }{ \mu^{\Phi^{!0}}_j } \big], 1 \big\} \Big]
\nonumber\\
& \approx  \mathbb{E}\Big[ \min\big\{ \big( 1 + \frac{ 1 }{ \mathcal{D}_{0j} } \big) \cdot \frac{ \xi }{ \mathbb{E}[ \mu^{\Phi^{!0}}_j ] }, 1 \big\} \Big].
\end{align}
In order to calculate the right hand side (R.H.S.) of \eqref{equ:Cnd_ActProb}, on the one hand, we use Slivnyark's theorem \cite{BacBla:09} and arrive at the following
\begin{align} \label{equ:Cnd_ps}
\mathbb{E} \big[ \mu^{\Phi^{!0}}_j \big] = \mathbb{E}^{!0} \big[ \mu^{\Phi}_j \big] = \mathbb{E} \big[ \mu^{\Phi}_j \big] = p_{ \mathrm s }.
\end{align}
On the other hand, by noticing that $\mathcal{D}_{ 0j } = \Vert X_0 - y_j \Vert^\alpha / T r^\alpha$ and $\Vert y_0 - y_j \Vert = u$, we can use the cosin law and express $\Vert X_0 - y_j \Vert$ as follows (see e.g., Fig.~\ref{fig:NodeLoc_v1}):
\begin{align} \label{equ:Dist_X0yj}
\Vert X_0 - y_j \Vert = \sqrt{ r^2 + u^2 - 2 u r \cos \Psi }
\end{align}
where $\Psi$ is the angle, which is a random variable, between the line segment connecting $X_0$ and $y_0$ and that of $y_j$ and $y_0$. Under the Poisson bipolar network, this quantity is uniformly distributed on $[0, 2\pi)$, with the probability density function (PDF) given as
\begin{align} \label{equ:f_psi}
f_\Psi(\psi) = \frac{1}{ 2 \pi }, ~~\psi \in [0, 2\pi).
\end{align}
The result in Lemma~\ref{lma:DistCnd_ActProb} immediately follows by taking \eqref{equ:Cnd_ps}, \eqref{equ:Dist_X0yj}, and \eqref{equ:f_psi} into \eqref{equ:Cnd_ActProb} and conduct algebraic computation.

\subsection{Proof of Theorem~\ref{thm:ST_SINR_CoverProb} } \label{apx:ST_SINR_CoverProb}
Given the typical link is active, we can take an expectation on both sides of \eqref{equ:CondThrPut} and arrive at the following
\begin{align} \label{equ:SucProb_temp}
& e^{ \frac{ \theta r^\alpha }{ \rho } } p_{ \mathrm s } = \mathbb{E}\Big[ \prod_{ j \neq 0 } \big( 1 - \frac{ a_j^\Phi }{ 1 + \mathcal{D}_{j0} } \big) \Big]
\nonumber\\
&\stackrel{(a)}{=} \mathbb{E}_{ \hat{\Phi} }^0\Big[ \prod_{ j \neq 0 } \mathbb{E}\big[ 1 - \frac{ a_j^\Phi }{ 1 + \mathcal{D}_{j0} } \big| \Vert y_j - y_0 \Vert = l \big]  \Big]
\nonumber\\
&\stackrel{(b)}{=} \mathbb{E}_{ \hat{\Phi} }\Big[ \prod_{ j \neq 0 } \big( 1 - \frac{ \mathbb{P}( \zeta_j = 1 \big| \Vert y_j - y_0 \Vert = l, \zeta_0 = 1 ) }{ 1 + \Vert \sqrt{ l^2 + r^2 - 2 r l \cos \Psi } \Vert^\alpha / \theta r^\alpha } \big) \Big]
\nonumber\\
&= \exp\!\Big(\! \!-\! \frac{ \lambda }{ 2 \pi } \! \int_{ \mathbb{R}^2 } \! \int_0^{ 2\pi } \!\!\! \frac{ \mathbb{P}( \zeta_x = 1 \big| \Vert x \Vert = l, \zeta_0 = 1 ) }{ 1 \!+\! { ({ l^2 \!+\! r^2 \!-\! 2 r l \cos \varphi })^{ \frac{ \alpha }{ 2 } } }\!/{ \theta r^\alpha } } d \varphi d x \Big),
\end{align}
where ($a$) is to take the expectation of the point process $\Phi$ by firstly conditioning with respect to the locations of receivers $\hat{\Phi}$, and ($b$) follows by applying the cosine law and the Slivnyark's theorem \cite{BacBla:09}. The expression in \eqref{equ:CovProb_exct} can then be attained by substituting \eqref{equ:CondAct_j} into \eqref{equ:SucProb_temp} and perform further algebraic manipulations.

Note that the equation \eqref{equ:CovProb_exct} involves two sets of integrals with respect to random angles, e.g., $\varphi$ and $\psi$ in Fig.~\ref{fig:NodeLoc_v1}, on $[0, 2\pi)$. This is because we are deconditioning the point process $\Phi$ with respect to $\hat{\Phi}$.
To that end, a simplified approximation of the SINR coverage probability, i.e., equation \eqref{equ:approx_CovProb}, can be attained by replacing the distances $\Vert X_j - y_0 \Vert$ and $\Vert X_0 - y_j \Vert$ by $\Vert y_j - y_0 \Vert$, which largely accelerates the computational efficiency.
\begin{figure}[t!]
  \centering{}

    {\includegraphics[width=0.95\columnwidth]{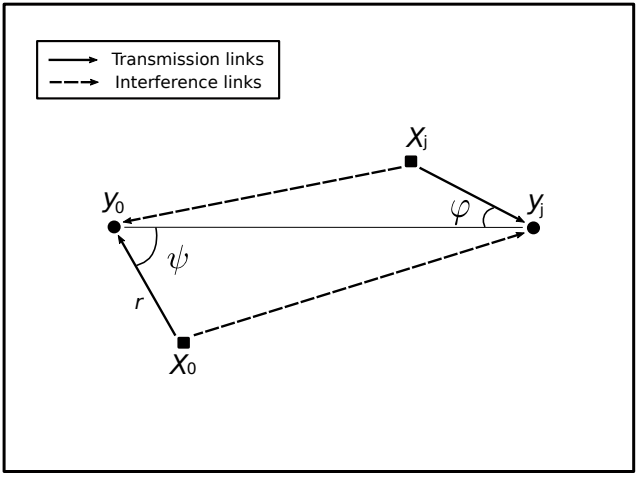}}

  \caption{ Example of a two-points location topology given the distance between receivers being $\Vert y_0 - y_j\Vert = u$. }
  \label{fig:NodeLoc_v1}
\end{figure}

\subsection{Proof of Theorem~\ref{thm:Meta_SINR}} \label{apx:Meta_SINR}
For ease of exposition, let us denote $\mathcal{F}_t$ as the $\sigma$-algebra that contains all the information about the queueing state of every link up to time slot $t$.
Note that in a queueing system, such $\sigma$-algebra forms a filtration, i.e., $\mathcal{F}_{t-1} \subset \mathcal{F}_t$.
We further introduce two parameters $Y_{0,t}^\Phi$ and $q_{u,t}$ whereas $Y_{0,t}^\Phi = \ln \mathbb{P}( \gamma_{0,t}^\Phi > \theta | \Phi )$ and $q_{u,t} = \mathbbm{1}\{ \zeta_{j,t} = 1 | \Vert y_j - y_0 \Vert = u, \zeta_{0,t} = 1 \}$, respectively.

At the initial state (i.e., $t=0$) of the queueing network, packets arrive at each node with probability $\xi$, and hence $\mathbb{E}[ q_{u,0} ] = \xi$. As such, the moment generating function of $Y_{0,0}^\Phi$ at the typical transmitter can be calculated as follows:
\begin{align} \label{equ:Intl_MmnGen_Y}
& M_{ Y^\Phi_{0,0} }(s) = \mathbb{E}\big[ \mathbb{P}( \gamma_{0,0} > \theta | \Phi )^s \big]
\nonumber\\
&= e^{ - \frac{ s \theta r^\alpha }{ \rho } } \mathbb{E} \bigg[ \prod_{ j \neq 0 } \!\Big(  1 - \frac{ \xi }{ 1  +  \Vert X_j \!-\! y_0 \Vert^\alpha \!/ \theta r^\alpha } \Big)^s \bigg]
\nonumber\\
&= \exp\!\left(\!{ - \frac{ s \theta r^\alpha }{ \rho } - \lambda \frac{ \delta \pi^2 r^2 \theta^\delta  }{ \sin( \pi \delta ) } \sum_{k=1}^\infty \!  \binom{ s }{ k } \! \binom{ \delta \!-\! 1 }{ k \!-\! 1 } (-1)^{k+1} \xi^k }\right).
\end{align}
We can then compute the CDF of $\mu_{0,0}^\Phi$ using the Gil-Pelaez theorem \cite{Gil}, as follows:
\begin{align} \label{equ:F_nu_0}
F_{\theta,0}(u) &= \mathbb{P}( \mathbb{P}( \gamma_{0,0} > \theta | \Phi ) < u ) = \mathbb{P}( Y^\Phi_{0,0} < \ln u )
\nonumber\\
& = \frac{1}{2} - \frac{1}{\pi} \! \int_{0}^{\infty} \!\!\!\! \mathrm{Im} \Big\{ u^{ - j \omega } M_{ Y^\Phi_{0,0} }(j \omega) \Big\} \frac{ d \omega }{ \omega }.
\end{align}

Next, let us consider the queueing system has involved to the $n$-th state, i.e., $t=n$. At this stage, the CDF of $\mu_{j,n-1}^\Phi$, namely $\mathbb{P}( \mu_{j,n-1}^\Phi < u ) = F_{\theta, n-1}(u)$, can be readily attained in an iterative manner. By leveraging Lemma~\ref{lma:CondThrPut}, we compute the moment generating function of $Y^\Phi_{0,n}$ as
\begin{align} \label{equ:MmnGen_Y}
& M_{ Y^\Phi_{0,n} }(s) = \mathbb{E}\big[ \mathbb{P}( \gamma_{0,n} > \theta | \Phi )^s \big]
\nonumber\\
=& e^{ - \frac{ s \theta r^\alpha }{ \rho } } \mathbb{E} \bigg[ \prod_{ j \neq 0 } \!\Big(  1 - \frac{ a_{j,n}^\Phi }{ 1  +  \Vert X_j \!-\! y_0 \Vert^\alpha \!/ \theta r^\alpha } \Big)^s \bigg]
\nonumber\\
\stackrel{(a)}{=} &e^{ - \frac{  s \theta r^\alpha }{ \rho } } \mathbb{E}^0_{ \hat{\Phi} } \! \bigg[\! \prod_{ j \neq 0 } \!\!\Big(\!  1 \!-\! \frac{ a_{j,n}^\Phi \theta }{ \theta  \!+\!  \vert  ( \Vert y_j \!-\! y_0 \Vert/r - \cos \! \Psi )^2 \!+\! \sin^2 \! \Psi \vert^{ \frac{\alpha}{2} }  } \!\Big)^s \bigg]
\nonumber\\
\stackrel{(b)}{=} & \exp\! \bigg( \! \frac{ - s \theta r^\alpha }{ \rho } \! -\! \! \int_0^\infty \!\!\!\! \int_0^{2\pi} \!\!\!\!  \frac{ \lambda \! \sum_{k=1}^{s} \!\! \binom{s}{ k } (-1)^{k+1} \mathbb{E}[q_{u,n}^k] d \varphi u du }{ \big[ 1 \!+\! ( r^2 \!+\! u^2 \!-\! 2 u r \cos \varphi )^{ \frac{\alpha}{2} }\!/\theta r^\alpha \big]^k } \bigg),
\end{align}
where ($a$) is by using the cosine law and ($b$) follows from applying the Slivnyark's theorem and taking expectation according to the point process $\hat{\Phi}$.
The complete expression of \eqref{equ:MmnGen_Y} requires us to calculate $\mathbb{E}[q_u^k]$, which can be written as follows:
\begin{align} \label{equ:E_q_k}
\mathbb{E}[q_{u,n}^k] = \mathbb{E}\bigg[ \Big( \frac{ \xi  \!+\! \xi /\mathcal{D}_{0j}  }{ \mu_{j,n}^{\Phi^{!o}} } \Big)^k \mathbbm{1}\Big\{ \mu_{j,n}^{\Phi^{!o}} \geq \xi \big( 1 + \frac{1}{ \mathcal{D}_{0j} } \big)\Big\}
\nonumber\\
+  \mathbbm{1}\Big\{ \mu_{j,n}^{\Phi^{!o}} < \xi \big( 1 + \frac{1}{ \mathcal{D}_{0j} } \big) \Big\} \Big| \Vert y_j - y_0 \Vert = u, \mathcal{F}_{n-1} \bigg].
\end{align}
As such, using the Slivnyark's theorem another time, the first term on the right hand side (R.H.S.) of \eqref{equ:E_q_k} can be computed as
\begin{align} \label{equ:Eq_term1}
&\mathbb{E}\bigg[ \Big( \frac{ \xi  \!+\! \xi/\mathcal{D}_{0j} }{ \mu_{j,n}^{\Phi^{!o}} } \Big)^k \!\!\! \mathbbm{1}\!\Big\{ \mu_{j,n}^{\Phi^{!o}} \!\! \geq\! \xi \big( 1 \!+\! \frac{1}{ \mathcal{D}_{0j} } \big) \!\Big\} \Big| \Vert y_j \!-\! y_0 \Vert \!=\! u, \mathcal{F}_{n-1}  \bigg]
\nonumber\\
&= \int_{0}^{ 2\pi } \!\!\! \int_{ \mathcal{H}_\theta( u/r, \psi ) }^{ 1 } \!\!\!\!\!\!\!\!\!\!\!\!\!\!\! \mathcal{H}^k_\theta( u/r, \psi ) \frac{ F_{ \theta,n-1 }(dt) }{ t^k }  \frac{ d \psi }{ 2\pi }.
\end{align}
Similarly, we can obtain the expression for the second term on the R.H.S. of \eqref{equ:E_q_k} as follows:
\begin{align} \label{equ:Eq_term2}
& \mathbb{E}\bigg[ \mathbbm{1}\Big\{ \mu_{j,n}^{\Phi^{!o}} \! <\! \xi \big( 1 \!+\! \frac{1}{ \mathcal{D}_{0j} } \big) \Big\} \Big| \Vert y_j \!-\! y_0 \Vert \!=\! u, \mathcal{F}_{n-1}  \bigg]
\nonumber\\
&= \int_{0}^{2\pi} \!\!\!\!\!  F_{ \theta,n-1 } \big( \mathcal{H}_\theta( u/r, \psi ) \big) \frac{ d \psi }{ 2\pi }.
\end{align}

Using the Gil-Pelaez theorem for another time, we have the CDF of $\mu_{0,n}^\Phi$ given as follows:
\begin{align} \label{equ:F_nu}
F_{\theta,n}(u) &= \mathbb{P}( \mathbb{P}( \gamma_{0,n} > \theta | \Phi ) < u ) = \mathbb{P}( Y^\Phi_{0,n} < \ln u )
\nonumber\\
& = \frac{1}{2} - \frac{1}{\pi} \! \int_{0}^{\infty} \!\!\!\! \mathrm{Im} \Big\{ u^{ - j \omega } M_{ Y^\Phi_{0,n} }(j \omega) \Big\} \frac{ d \omega }{ \omega }.
\end{align}
Note that $F_{\theta,n}(u)$ appears on the left hand side of \eqref{equ:F_nu}, and $F_{\theta, n-1}(\cdot)$ is implicitly contained in the right hand side of \eqref{equ:F_nu}. Because $\forall u \in [0,1]$ and $n\in \mathbb{N}$, it holds that $F_{\theta,n}(u) \leq F_{\theta,0}(u)$, by the Dominated Convergence Theorem, we have $F_{\theta,n}(u) \rightarrow F_\theta(u)$ as $n \rightarrow \infty$.
To this end, by substituting \eqref{equ:Eq_term1}, \eqref{equ:Eq_term2}, and \eqref{equ:MmnGen_Y} into \eqref{equ:F_nu} and taking $n \rightarrow \infty$, we have the desired result.

\end{appendix}

\bibliographystyle{IEEEtran}
\bibliography{bib/StringDefinitions,bib/IEEEabrv,bib/howard_ItrcQue}

\end{document}